\documentclass[fleqn,twoside]{article}
\usepackage{amsmath,amssymb,espcrc2,color,cite}

\usepackage{graphicx,soul}
\usepackage[figuresright]{rotating}

\newcommand{\url}[1]{{\tt #1}}
\newcommand{\lsim}
{\;\raisebox{-.3em}{$\stackrel{\displaystyle <}{\sim}$}\;}
\newcommand{\gsim}
{\;\raisebox{-.3em}{$\stackrel{\displaystyle >}{\sim}$}\;}

\newcommand{\gmt}{\ensuremath{(g-2)_\mu}}
\newcommand{\br}{{\rm BR}}
\newcommand{\bsg}{BR($b \to s \gamma$)}
\newcommand{\btn}{BR($B_u \to \tau \nu_\tau$)}

\newcommand{\bmm}{\ensuremath{\br(B_s \to \mu^+\mu^-)}}

\newcommand{\rmm}{\ensuremath{R_{\mu \mu}}}
\newcommand{\ssi}{\ensuremath{\sigma^{\rm SI}_p}}

\newcommand{\Och}{\ensuremath{\Omega_\chi h^2}}

\newcommand{\Mh}{M_h}

\newcommand{\MA}{M_A}

\newcommand{\mgl}{m_{\tilde g}}
\newcommand{\msq}{m_{\tilde q}}
\newcommand{\msqr}{m_{\tilde q_R}}
\newcommand{\cha}[1]{\tilde \chi^\pm_{#1}}

\newcommand{\mcha}[1]{m_{\tilde \chi^\pm_{#1}}}
\newcommand{\neu}[1]{\tilde \chi^0_{#1}}
\newcommand{\mneu}[1]{m_{\tilde \chi^0_{#1}}}

\newcommand{\tb}{\tan\beta}

\newcommand{\tev}{\,\, \mathrm{TeV}}
\newcommand{\gev}{\,\, \mathrm{GeV}}

\def\reffi#1{\mbox{Fig.~\ref{#1}}}

\definecolor{orange}{rgb}{1,0.5,0}

\newcommand{\ETslash}{/ \hspace{-.7em} E_T}

\graphicspath{{figs/}}

\title{\vspace{-1.5cm}
\bf The NUHM2 after LHC Run 1 \\ \vspace{0.5em}}

\author{
{\bf O.~Buchmueller}\address[Imperial]
   {High\,Energy\,Physics\,Group,\,Blackett\,Laboratory,\,Imperial\,College,\,Prince\,Consort\,Road,\,London\,SW7\,2AZ,\,UK},
{\bf R.~Cavanaugh}\address[FNAL]
   {Fermi National Accelerator Laboratory, P.O. Box 500, 
    Batavia, Illinois 60510, USA}\hbox{$^{\rm ,}$}\address[UIC]
   {Physics Department, University of Illinois at Chicago, Chicago, 
    Illinois 60607-7059, USA},
{\bf M.~Citron}\addressmark[Imperial],
{\bf A.~De Roeck}\address[CERN]
   {Physics Department, CERN, CH--1211 Gen\`eve 23, Switzerland}\hbox{$^{\rm ,}$}\address[Antwerpen]
   {Antwerp University, B--2610 Wilrijk, Belgium},
 {\bf M.J.~Dolan}\address[SLAC]
{Theory Group, SLAC National Accelerator Laboratory,
2575 Sand Hill Road, Menlo Park, \\ CA 94025-7090, USA},
{\bf J.R.~Ellis}\address[KCL]{Theoretical Particle Physics
  and Cosmology Group, Department of Physics, King's College London, London~WC2R~2LS, UK}\hbox{$^{\rm ,}$}\addressmark[CERN], 
{\bf H.~Fl\"acher}\address[Bristol]
   {H.H.~Wills Physics Laboratory, University of Bristol, Tyndall Avenue, Bristol BS8 1TL, UK},
{\bf S.~Heinemeyer}\address[Santander]
   {Instituto de F\'{\i}sica de Cantabria (CSIC-UC), 
    E--39005 Santander, Spain},
{\bf S.~Malik}\addressmark[Imperial],
{\bf J.~Marrouche}\addressmark[CERN],
{\bf D.~Mart\'inez~Santos}\address[NIKHEF]{NIKHEF\,and\,VU\,University\,Amsterdam,
Science\,Park\,105,\,NL-1098\,XG\,Amsterdam,\,The\,Netherlands},
{\bf K.A.~Olive}\address[Minnesota] 
{William I.\ Fine Theoretical Physics Institute, School of Physics and
 Astronomy, University of Minnesota, Minneapolis, Minnesota 55455, USA}, 
{\bf K.J.~de~Vries}\addressmark[Imperial],
{\bf G.~Weiglein}\address[DESY]
   {DESY, Notkestra{\ss}e 85, D--22607 Hamburg, Germany}
}

\begin{document}

\begin{abstract}
We make a frequentist analysis of the parameter space of the NUHM2,
in which the soft supersymmetry (SUSY)-breaking contributions to
the masses of 
the two Higgs multiplets, $m^2_{H_{u,d}}$, vary independently from the
universal soft SUSY-breaking contributions $m^2_0$ to the masses
of squarks and sleptons. Our analysis uses the {\tt MultiNest} sampling algorithm
with over $4 \times 10^8$ points to sample the NUHM2 parameter space.
It includes the ATLAS and CMS Higgs mass measurements
as well as their searches for supersymmetric jets + $\ETslash$ 
signals using the full LHC Run~1 data, the measurements of \bmm\ by LHCb and CMS
together with other B-physics observables, electroweak precision observables 
and the XENON100 and LUX searches for spin-independent dark matter scattering.
We find that the preferred regions of the NUHM2 parameter space have
negative SUSY-breaking scalar masses squared for squarks and sleptons, $m_0^2 < 0$,
as well as $m^2_{H_u} < m^2_{H_d} < 0$.
The tension present in the CMSSM and NUHM1 between the supersymmetric
interpretation of \gmt\ 
and the absence to date of SUSY at the LHC
is not significantly alleviated in the NUHM2. We find that the minimum $\chi^2 = 32.5$ with 21
degrees of freedom (dof) in the NUHM2, to be compared with 
$\chi^2/{\rm dof} = 35.0/23$
in the CMSSM, and $\chi^2/{\rm dof} = 32.7/22$ in the NUHM1.
We find that the one-dimensional
likelihood functions for sparticle masses and other observables are similar to those found
previously in the CMSSM and NUHM1.

\vspace{0.5em}
\begin{center}
{\tt KCL-PH-TH/2014-33, LCTS/2014-29, CERN-PH-TH/2014-145, \\
DESY 14-144, FTPI-MINN-14/39, UMN-TH-3344/14, SLAC-PUB-16051}
\end{center}

\vspace{2.0cm}
\end{abstract}

\maketitle

\section{Introduction}
\label{sec:intro}

Supersymmetric (SUSY) models are among the best-motivated extensions of the Standard Model (SM)
that might be discovered at the Large Hadron Collider (LHC). They stabilize the electroweak hierarchy~\cite{Ellis:2000ig} and
facilitate grand unification~\cite{unify}, and the lightest supersymmetric particle (LSP) provides a natural candidate for the 
cosmological dark matter~\cite{EHNOS}. However, the absence of a signal in direct searches for SUSY
particles in Run~1 of the LHC~\cite{ATLAS20,CMS20} sets strong constraints on supersymmetric models,
as do the measurement of the mass and 
properties of the Higgs boson~\cite{lhch} and precision measurements of rare decays such as 
$B_s \to \mu^+ \mu^-$~\cite{LHCbBsmm,CMSBsmm,BsmmComb} . 

Gaining a fully accurate picture of the effects of these constraints requires that they be
combined in global statistical fits within specific supersymmetric models. 
Particularly well-motivated and simplified versions of the minimal supersymmetric Standard Model 
(MSSM)~\cite{HK} are derived from grand unified theory (GUT) model-building considerations. 
There have been a number of analyses~\cite{mc9,otherRun1,previousNUHM2} 
of the constraints imposed by LHC Run~1 data on the parameter spaces of such models, 
particularly the constrained MSSM (CMSSM)~\cite{funnel,cmssm,AbdusSalam:2011fc},
whose parameters are the soft supersymmetry (SUSY)-breaking masses $m_0$, $m_{1/2}$ and $A_0$
that are universal at the GUT scale, and $\tb$, the ratio of the two vacuum expectation values of the two Higgs doublets.
There have also been some studies of the LHC constraints on the NUHM1~\cite{nuhm1}, in which the soft SUSY-breaking contributions to the masses of the electroweak Higgs multiplets, $m^2_{H_d, H_u}$, are equal but non-universal. 

However, these models have become very constrained by the recent data. 
The anomalous magnetic moment of the muon \gmt~\cite{newBNL,g-2} is a particular source of tension,
as has been reinforced by the recent convergence in the Standard Model (SM) calculations of 
\gmt\ based on $\tau$ decays and  different sets of $e^+ e^-$ annihilation data~\cite{Jegerlehner}.
As is well known, the $\sim 3.5\,\sigma$ discrepancy between the observed value and SM prediction
can be reduced by SUSY contributions due to relatively light electroweakly-interacting superpartners. 
In the simple GUT-based models mentioned above, direct searches and the Higgs mass force the
coloured super-partners to be so heavy that, due to the universality of the soft SUSY-breaking parameters 
$m_0$ and $m_{1/2}$ at the GUT scale that leads also to relatively heavy electroweak superpartners,
these models cannot remove the \gmt~ anomaly~\cite{mc7}. 

A related extension of these models which \textit{a priori} might be able to alleviate this tension is the
NUHM2~\cite{nuhm2}, in which $m^2_{H_d} \ne m^2_{H_u} \ne m_0^2$ in general~\footnote{For previous studies of the 
NUHM2 in light of LHC data, see~\cite{previousNUHM2}.}, but the soft SUSY-breaking parameters 
$m_0$, $m_{1/2}$ and $A_0$ are still universal at the GUT scale. An equivalent formulation of the NUHM2 
is to treat the pseudoscalar mass $\MA$ and supersymmetric Higgs mass term $\mu$ as free parameters,
which could lead to interesting phenomenology associated with light
higgsinos and/or a light pseudoscalar Higgs.
Moreover, new terms in the renormalization group equations (RGEs) associated
with the scalar-mass 
non-universality in the NUHM2 may lead to lighter left-handed sleptons, offering further avenues for ameliorating the tension with 
\gmt\ (see Sec.~\ref{sec:sparam} for details).

Therefore, in this paper we extend our previous analyses of the CMSSM and 
NUHM1~\cite{mc9} to the NUHM2~\cite{nuhm2}, and compare the corresponding phenomenological predictions. 
In addition to the 8~TeV ATLAS search for supersymmetry in the jets + $\ETslash$~\cite{ATLAS20}~\footnote{See
also~\cite{CMS20}, which we do not use in our analysis.} 
channel, our frequentist fit using the {\tt MultiNest} ~\cite{Feroz:2008xx} sampling algorithm includes 
Higgs mass measurements~\cite{lhch}, the measurements of \bmm\ by LHCb and 
CMS~\cite{LHCbBsmm,CMSBsmm,BsmmComb}, other B-physics~\cite{HFAG} and electroweak precision observables~\cite{EWWG}, 
and the XENON100 and LUX searches for spin-independent dark matter scattering~\cite{XENON100,lux}.

We find that the NUHM2, despite its freedom in the choices of $\MA$ and $\mu$,
is unable to alleviate significantly the tension between the absence to date of SUSY at the LHC
and the supersymmetric interpretation of \gmt\ that had been found previously in the CMSSM and NUHM1.
We find that the minimum  $\chi^2/{\rm dof} = 32.5/21$ in the NUHM2, 
to be compared with $\chi^2/{\rm dof} = 35.0/23$ in the CMSSM and $\chi^2/{\rm dof} = 32.7/22$ in the NUHM1.
A novel feature of the best NUHM2 fit is that the preferred regions of the NUHM2 parameter space
have negative SUSY-breaking scalar masses squared for squarks and sleptons, $m_0^2 < 0$,
as well as $m^2_{H_u} < m^2_{H_d} < 0$.%
\footnote{Negative SUSY-breaking scalar masses have also 
arisen in recent post-Higgs gauge mediation constructions~\cite{Craig:2012xp}. For a discussion of cosmological issues
associated with such tachyonic soft SUSY-breaking mass parameters,
see~\cite{EGLOS}.}

As an output of our analysis, we compare the one-dimensional
likelihood functions for sparticle masses and other observables in the NUHM2 with those found
previously in the CMSSM and NUHM1. The 95\% CL lower limits on the gluino,
squark, stop and stau masses are not very different in the NUHM2 from those found
previously in the CMSSM and NUHM1. However, the distinction found in those
models between low- and high-mass regions of their respective parameter spaces is
largely lost in the NUHM2 because of its greater flexibility in satisfying the dark
matter constraint. In addition to sparticle masses, we also present NUHM2 predictions
for \bmm\ and the spin-independent dark-matter scattering cross section,~\ssi.


\section{Analysis Procedure}

We follow closely the procedure described in~\cite{mc9}.
Our treatment of the non-LHC constraints is identical with the treatment
in~\cite{mc9} (minor shifts in some observables, such as in the top
quark mass, do not have a relevant impact), and we again use the 
{\tt MultiNest} algorithm 
to sample the NUHM2 parameter space, just as we did previously for
the CMSSM and NUHM1 models. As mentioned in the Introduction, we use
a NUHM2 sample comprising $\sim 4 \times 10^8$ points, with the aim of sampling
adequately features of the six-dimensional NUHM2 parameter
space $\{m_0, m_{1/2}, m_{H_u}, m_{H_d}, A_0, \tb \}$, ensuring in particular
that all high-likelihood regions are identified and well characterized.
We sample the ranges $-1333 \gev < m_0 < 4000 \gev$, $0 < m_{1/2} < 4000 \gev$,
$- 5\times 10^7 \gev^2 < m_{H_u}^2, m_{H_d}^2 < 5\times 10^7 \gev^2$,
$- 8000 \gev < A_0 < 8000 \gev$ and $2 < \tb < 68$. 
(Here and subsequently, negative values of $m_0$ should be
understood as $m_0 \equiv {\rm Sign}(m_0^2) \sqrt{|m_0^2|} < 0$, 
and we use analogous definitions for negative values of $m_{H_u}$ and $m_{H_d}$.)
The parameter ranges are scanned by
dividing the range of $m_0$ into 4 segments, and the ranges of 
$m_{1/2}, m_{H_u}$ and $m_{H_d}$ into 3 segments each, 
yielding a total of 108 boxes. Their boundaries are smeared using a
Gaussian function
so as to sample the NUHM2 parameter space smoothly, which also
provides some information beyond the nominal sampling range, as we see later
in the case of $m_{H_u}$ and $m_{H_d}$.

We merge this dedicated sample of the NUHM2 parameter space with
the samples of the CMSSM and NUHM1 parameter spaces used in~\cite{mc9}.
The latter are subspaces of the full NUHM2 parameter space, and the CMSSM and
NUHM1 points provide supplementary sampling of the likelihood function of the NUHM2.

We construct a global likelihood function that receives contributions
from the usual electroweak precision observables, as well as B-decay
measurements such as \bsg, \btn\ and \bmm.
Bounds on their experimental values as well as those on the 
cosmological dark matter density, the cross-section for spin-independent
dark matter scattering from the LUX experiment and the LHC searches for
supersymmetric signals are given in~\cite{mc7}, with updates detailed in
\cite{mc8}. 
Their contributions to the likelihood function
are calculated within the 
{\tt MasterCode} framework~\cite{mcweb}. 
This incorporates a code for the electroweak
observables based on~\cite{Svenetal} as well as the {\tt SoftSUSY}~\cite{Allanach:2001kg}, {\tt FeynHiggs}~\cite{FeynHiggs,Mh-logresum}, 
{\tt SuFla}~\cite{SuFla}, {\tt SuperIso}~\cite{SuperIso}, 
{\tt MicrOMEGAs}~\cite{MicroMegas}  
and {\tt SSARD}~\cite{SSARD} codes, using the SUSY Les Houches
Accord~\cite{SLHA}.
The ATLAS and CMS measurements of the Higgs mass, $\Mh$, are interpreted
using {\tt FeynHiggs~2.10.0}~\cite{Mh-logresum} to calculate $\Mh$ and 
as in \cite{mc9} we allow conservatively for a theoretical
uncertainty of 1.5~GeV at each point in the NUHM2 parameter space~%
\footnote{As in~\cite{mc9}, we do not include constraints
from the Higgs signal strength measurements. These are not yet sufficiently accurate to constrain our results,
since the Higgs rate predictions in
the favoured regions of the NUHM2, NUHM1 and CMSSM parameter space
lie in the decoupling regime, despite the additional freedom for
$\MA$ in the NUHM2 and NUHM1.}.%
~The improvements recently
incorporated into {\tt FeynHiggs}\cite{FeynHiggs} yield
an upward shift of $\Mh$ for scalar top masses in the (multi-)TeV range
and reduce the theoretical uncertainty in the Higgs mass
calculation~\cite{mc8.5}.

We incorporate here the public results of searches for
jets + $\ETslash$ events without leptons using the full ATLAS Run~1 data set of $\sim 20$/fb at 8~TeV~\cite{ATLAS20}, which
has greater sensitivity to the relevant parts of the NUHM2 parameter space than searches including leptons.
The experimental searches for jets + $\ETslash$ events are typically
analyzed within the framework of the CMSSM for some fixed $A_0$ and
$\tb$. The applicability of these analyses to other $A_0$
and $\tb$ values, 
as well as to constraining the NUHM1,2, requires some study and
justification. One issue is that, for any specific set of values of
$m_0$, $m_{1/2}$, $A_0$ and $\tb$, the sensitivities of ATLAS and CMS to
jets + $\ETslash$ events might depend on the degree
of non-universality in the NUHM1,2. A second issue is that the range of
$m_0$ in the NUHM2 that is consistent with the $\neu{1}$ LSP requirement
depends on the degrees of non-universality. Specifically, this
requirement is compatible with $m^2_0 < 0$ in the NUHM2, a possibility
that is absent for the CMSSM, but can occur in the NUHM1
for $m_{1/2} \gsim 2000\gev$ when $m^2_{H_d} = m^2_{H_u} < 0$ and dominates over $m_0^2$
in the renormalization group evolution. In the NUHM2 it is even easier to obtain $m^2_0 < 0$ and remain
compatible with a neutralino LSP, because a
combination of soft supersymmetry-breaking parameters known as $S$ (defined below) may be non-zero.

Since the LHC experiments quote
limits only for the CMSSM with $m^2_0 > 0$, we rely on a previous
dedicated study of jets + $\ETslash$ searches at 7~TeV~\cite{mc8} made using the {\tt Delphes}~\cite{Delphes} 
generic simulation package with a `card' to emulate
the performance of the ATLAS detector, that showed that the LHC results could be extrapolated to $m_0^2 < 0$~%
\footnote{
CMSSM models with $m_0^2 < 0$ were studied in the context of a gravitino LSP
\cite{m2neg}. We recall that these and NUHM models with $m^2_{H_u} < 0$ and $m^2_{H_d} < 0$
are in principle subject to additional cosmological constraints \cite{EGLOS}.}.
This study confirmed that ~$\ETslash$ constraints in the $(m_0, m_{1/2})$ plane of the CMSSM
are relatively insensitive to $\tb$ and $A_0$,  as stated
in~\cite{cms0l-aT}, and that the~ $\ETslash$ constraints are also
quite insensitive to the degrees of non-universality in the NUHM1,2.
Specifically, it was found that the 95\% CL bounds in the 
$(m_0, m_{1/2})$ plane of the CMSSM were essentially independent of $A_0$ and $\tb$, 
as also stated by ATLAS~\cite{ATLASindependent},
that the same was true for $m_{H_u}^2 = m_{H_d}^2 \ne m_0^2$ 
in the NUHM1, and also for
$m_{H_u}^2 \ne m_{H_d}^2 \ne m_0^2$ in the NUHM2. The same is expected to be true for the 8-TeV ATLAS
jets + $\ETslash$ search~\cite{ATLAS20} used here, which uses a similar event selection.

Finally, we also incorporate here the most recent constraints on
$A/H$ production from ATLAS and CMS~\cite{ATLASHA}, using the
same approach as in~\cite{mc9}.

\section{Analysis of the NUHM2 Parameter Space}

\subsection{Scalar Mass Parameters and Renormalization}
\label{sec:sparam}

As is well known, in the CMSSM the electroweak vacuum conditions
may be used to determine both $\MA$ and $\mu$ for any fixed values of
$m_0, m_{1/2}, A_0$ and $\tb$, whereas in the NUHM1 the flexibility in $m_{H_u}^2 = m_{H_d}^2 \ne m_0^2$
permits one to treat either $\MA$ or $\mu$ as a free parameter, and in the NUHM2 the double flexibility in
$m_{H_u}^2 \ne m_{H_d}^2 \ne m_0^2$ allows both $\MA$ and $\mu$ to be treated as free parameters.

Before discussing our results for the NUHM2, we briefly review another important difference
between this model and its more constrained relatives. When $m_{H_u}^2 \ne m_{H_d}^2$,
the quantity~\cite{Martin:1993zk}
\begin{eqnarray}
S &\equiv& \frac{g_1^2}{4} \big( m_{H_u}^2 - m_{H_d}^2 +
        2 \big( m_{\widetilde{Q}_L}^2  - m_{\widetilde{L}_L}^2  \nonumber \\ && \, - \, 
	2 m_{\widetilde{u}_R}^2 +
	m_{\widetilde{d}_R}^2 + m_{\widetilde{e}_R}^2 \big) \nonumber \\ && \, + \, 
          \big( m_{\widetilde{Q}_{3L}}^2 - m_{\widetilde{L}_{3L}}^2 - 2
	  m_{\widetilde{t}_R}^2  \nonumber \\ && \, + \, 
	   m_{\widetilde{b}_R}^2 + m_{\widetilde{\tau}_R}^2 \big)\big) 
\label{defS}
\end{eqnarray}
is non-zero. In both the CMSSM and NUHM1, $S=0$ and is a fixed point of the RGEs at the 
one-loop level and remains zero at any
scale~\cite{Sterms}. 
However, in the NUHM2, with $m^2_{H_u} \ne m^2_{H_d}$,
$S \ne 0$ at the GUT scale, as seen in (\ref{defS}), which
can cause the low-energy spectrum to differ
significantly from that in the CMSSM or NUHM1.  For example, consider the renormalization-group equation for
the $\tau_R$ mass:
\begin{eqnarray}
 \frac{d m_{\widetilde{\tau}_R}^2}{dt} & =  & \frac{1}{8 \pi^2} (-4 g_1^2 M_1^2  \nonumber \\ && \, + \, 
          2 h_\tau^2 ( m_{\widetilde{L}_{3L}}^2 + m_{\widetilde{\tau}_R}^2 +
	  m_1^2 +  A_\tau^2 ) \nonumber  \\ && \, + \, 
	   4 S) \, .
\end{eqnarray}
When $S<0$, the evolution of $m^2_{\tau_R}$ receives a positive contribution as it runs 
down from the GUT scale to the electroweak scale.  As a result, ensuring a neutralino LSP
becomes a generic possibility even when $m^2_0 < 0$~\footnote{In the NUHM1,
the flexibility to allow $m^2_{H_d} = m^2_{H_u} < 0$ with a different value
from $m^2_0$ can also affect the running to ensure a neutralino LSP when $m^2_0 < 0$,
but only in a restricted region when $m_{1/2}$ is large enough.}.  Furthermore, the masses of left-handed sleptons may
run to lighter values than their right-handed counterparts, allowing for new co-annihilation 
channels to regulate the neutralino relic density \cite{nuhm2}, or larger contributions to~ \gmt.

\subsection{Model Parameter Planes}
\begin{figure*}[htb!]
\vspace{-0.7cm}
\begin{center}
\resizebox{7cm}{!}{\includegraphics{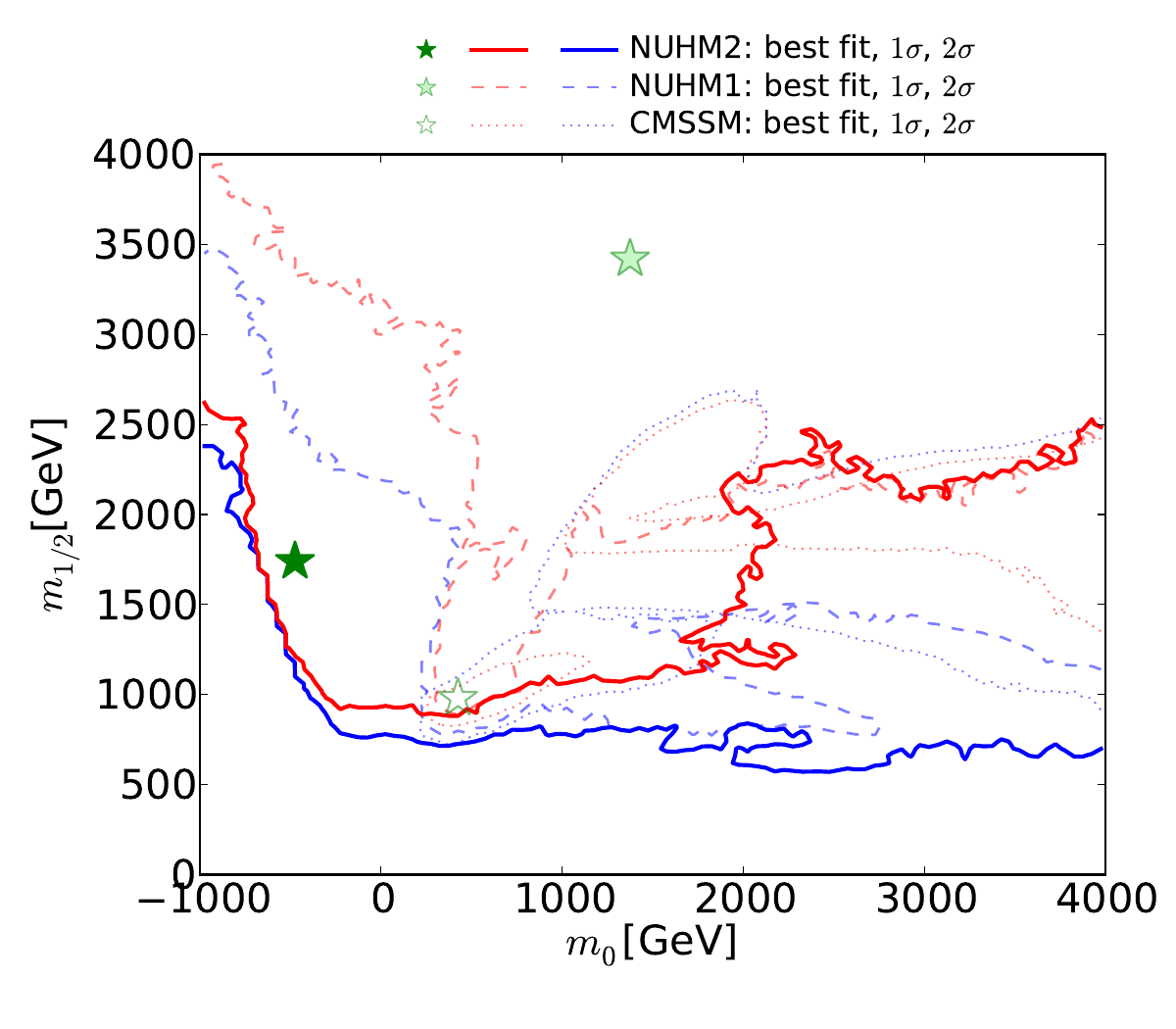}}
\resizebox{7.5cm}{!}{\includegraphics{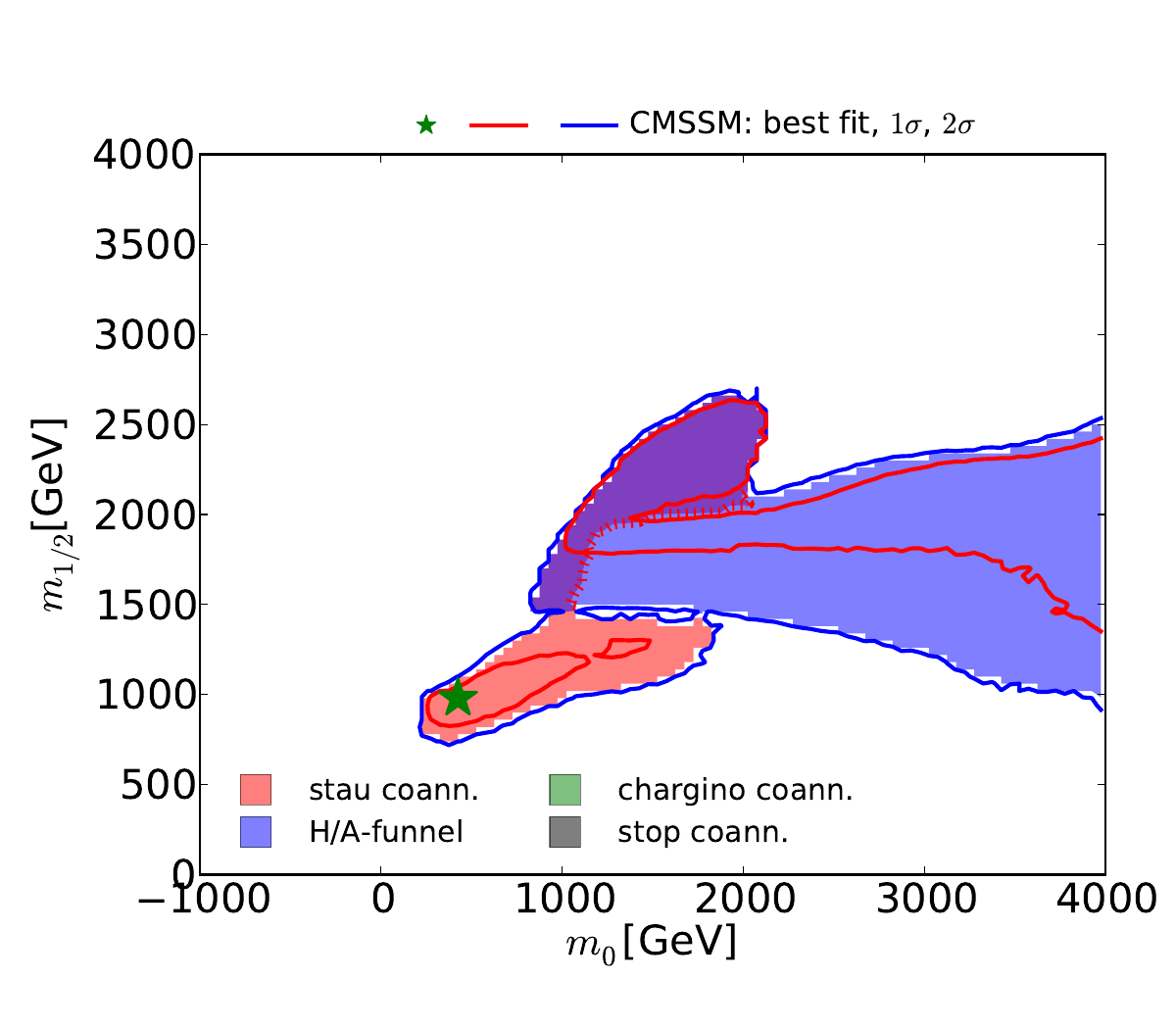}}
\resizebox{7.5cm}{!}{\includegraphics{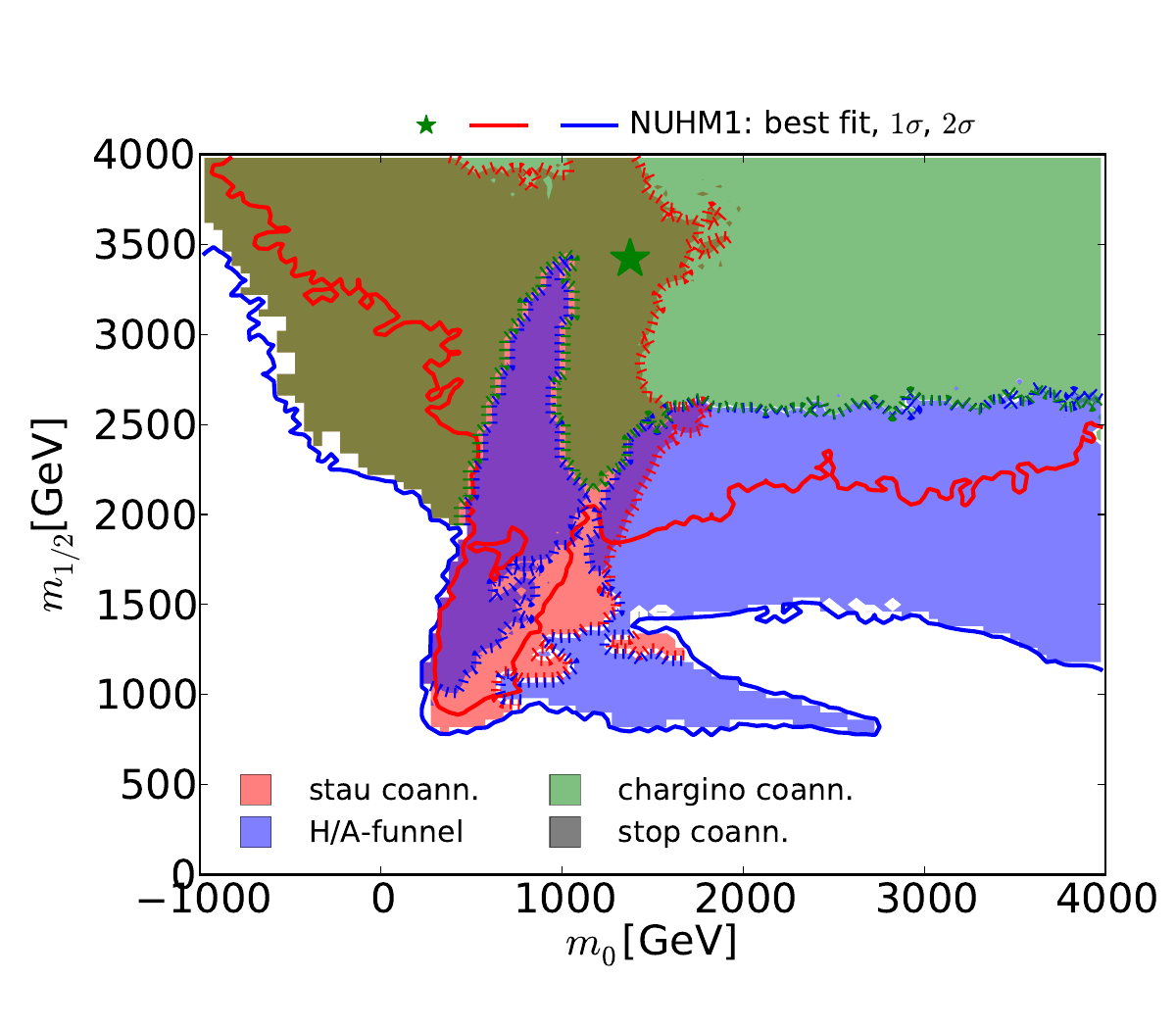}}
\resizebox{7.5cm}{!}{\includegraphics{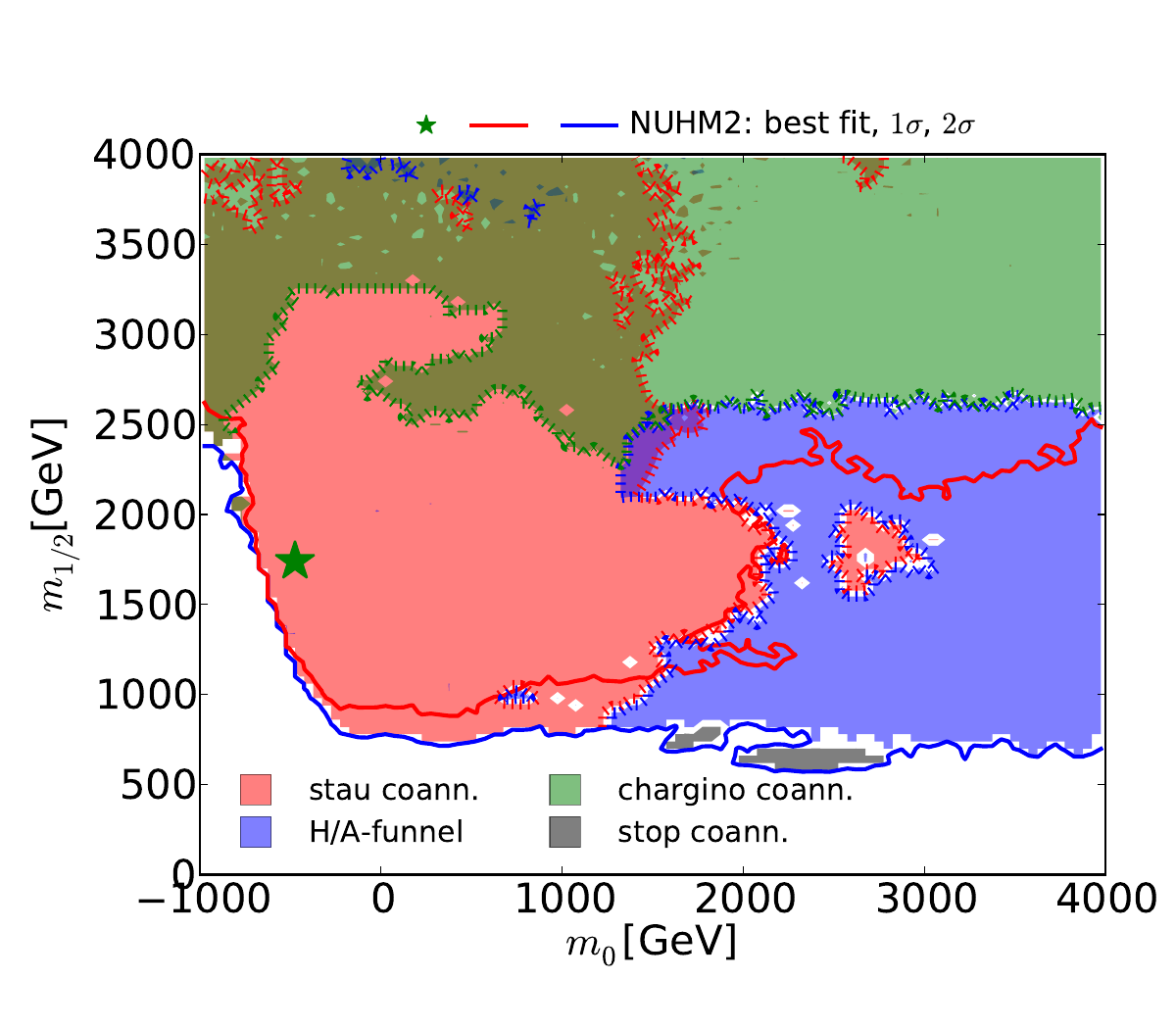}}
\end{center}
\vspace{-0.3cm}
\caption{\it Upper left: The $(m_0, m_{1/2})$ planes in the NUHM2, CMSSM and NUHM1. The
  results of the fit in the NUHM2 are indicated by solid lines and filled green
  stars, and those of our previous fits to the CMSSM and NUHM1 by dotted and
  dashed lines as well as open and shaded green stars,
  respectively. In all cases, the red lines denote  
  $\Delta \chi^2 = 2.30$ ($\sim 68$\% CL) contours, and the blue lines denote
  $\Delta \chi^2 = 5.99$ ($\sim 95$\% CL) contours. Upper right: The dominant mechanisms (\protect\ref{shadings}) fixing the
  dark matter density $\Och$ in the CMSSM. Lower left: The same for the NUHM1. Lower right: the same
  for the NUHM2. Stau coannihilation regions are shaded pink, rapid $A/H$ annihilation funnel regions are
  shaded blue, $\cha{1}$ coannihilation regions are shaded green, stop coannihilation regions are shaded grey.
  Regions where more than one of these conditions are satisfied are shaded in darker colours.}
\label{fig:m0m12}
\end{figure*}

~\\
\noindent
{\it The $(m_0, m_{1/2})$ plane:}\\
We first present results for the $(m_0, m_{1/2})$ plane shown in \reffi{fig:m0m12}.
We denote the best-fit point by a filled green star and the $\Delta \chi^2 = 2.30$
and 5.99 contours, corresponding approximately to the 68 and 95\% CL
contours, by solid red and blue contours, respectively. 
In the upper left panel of Fig.~\ref{fig:m0m12} we also show the best-fit points in the NUHM1 and CMSSM
(shaded and open green stars),
and the 68\% and 95\% CL contours in these models
(dashed and dotted red and blue contours, respectively).

We see that the 68\% CL NUHM2 region in the upper left panel of Fig.~\ref{fig:m0m12} extends
in a lobe down to $m_{1/2} \sim 1000 \gev$ for
$-500 \gev \lsim m_0 \lsim 2000 \gev$, whereas $m_0$ is relatively unrestricted for $m_{1/2} \gsim 2500 \gev$.
At the 95\% CL we find $m_{1/2} \gsim 500 \gev$ for $m_0 \gsim 0$. The
best-fit point in the NUHM2 has $m_0 \sim - 500 \gev$ and $m_{1/2} \sim 1800 \gev$. The LHC ~$\ETslash$
search with the most impact on the parameter space is that with jets and zero leptons, which constrains the NUHM2 parameter
space when $m_0 \lsim 1000 \gev$. As already mentioned,
we have verified previously~\cite{mc8} that this constraint is
essentially independent of the other NUHM2 parameters in the $(m_0, m_{1/2})$ region of
interest. Searches for events with $b$-jets and/or leptons have greater
sensitivity when $m_0 \gsim 1500 \gev$, but are important only outside the 95\% CL contour, at
lower $m_{1/2}$, so we have not studied in detail their sensitivity to the model parameters.

In the case of the NUHM1, the range of $m_0$ where low values of 
$m_{1/2} \lsim 2000 \gev$ are allowed at the 68\% CL (within the dashed red contour in Fig.~\ref{fig:m0m12}) is much
smaller, being limited to $200 \gev \lsim m_0 \lsim 1000 \gev$. The case of the CMSSM is
much more restrictive, with only a small part of the 68\% CL region (within the dotted red contour in Fig.~\ref{fig:m0m12})
with $300 \gev \lsim m_0 \lsim 1500 \gev$ appearing when $m_{1/2} \lsim 1800 \gev$. Moreover,
in this case at the 95\% CL the largest allowed value of $m_{1/2} \sim 2500 \gev$, whereas we observe
no upper bound on $m_{1/2}$ in either the NUHM1 or the NUHM2.

~\\
\noindent
{\it The dark matter constraint:}\\
The dark matter density constraint is less restrictive in the NUHM2
than in the NUHM1 and, particularly, the CMSSM.
In the regions of interest, the dark matter density is generally brought
down into the range allowed by cosmology through enhancement of
(co-)annihilation processes due to particular properties of the spectrum.
In the other panels of Fig.~\ref{fig:m0m12} we use different colours of
shading to visualize the impacts of these processes, by displaying
areas of the 95\% CL regions in the $(m_0, m_{1/2})$ planes where the following conditions are satisfied:
\begin{eqnarray}
{\tilde \tau_1} {\rm ~coannihilation~(pink):} & \frac{m_{\tilde \tau_1}}{\mneu{1}} - 1  & < \; 0.15 \, , \nonumber \\
A/H {\rm ~funnel~(blue):} & |\frac{\MA}{2\mneu{1}} - 1 | & < \; 0.2 \, , \nonumber \\
\cha{1} {\rm ~coannihilation~(green):} & \frac{\mcha{1}}{\mneu{1}} - 1  & < \; 0.1 \, , \nonumber \\
{\tilde t_1} {\rm ~coannihilation~(grey):} & \frac{m_{\tilde t_1}}{\mneu{1}} - 1  & < \; 0.2 \, . \nonumber
\label{shadings}
\end{eqnarray}
each of which is surrounded by a dotted contour. Regions where more than one of these conditions 
are satisfied are shaded in darker colours. We have also explored the focus-point criterion
$| \mu/\mneu{1} - 1| < 0.3$, and found that it is not relevant in the displayed portions of the $(m_0, m_{1/2})$ planes.
We note that the criteria above are approximate, being intended only to serve as guides to the different
regions in the $(m_0,m_{1/2})$ planes.

We see in the upper right panel of Fig.~\ref{fig:m0m12} that the low-mass
region of the CMSSM is in the stau coannihilation region \cite{stau-co,edsjo} 
(pink shading) and its high-$m_0$ region (blue shading) is in the 
funnel region where the LSPs annihilate rapidly through the s-channel heavy Higgs resonances $A/H$~\cite{funnel}.
The best-fit CMSSM point now lies in the stau coannihilation region: the difference from the
low-mass best-fit point found in~\cite{mc9}
is due to using the updated ATLAS jets + $\ETslash$ constraint~\cite{ATLAS20}. The current CMSSM best-fit
point is very similar to the previous local best fit in the low-mass region. 
We also see for $1000 \gev \lsim m_0 \lsim 2000 \gev$
and $m_{1/2} \gsim 2000 \gev$ (shaded purple) a CMSSM region where both the stau-coannihilation
and funnel criteria are satisfied.

In the NUHM1, as seen in the lower left panel of Fig.~\ref{fig:m0m12} it is possible to satisfy the $\Och$
constraint for larger values of $m_{1/2}$ than are possible in the CMSSM, thanks to the extra
degree of freedom associated with the soft SUSY-breaking contribution to the
Higgs masses. In the low-mass NUHM1 region, the relic density is again determined by stau coannihilation (pink shading),
whereas at large $m_0$ and $m_{1/2} \lsim 2500 \gev$ the rapid annihilation via the $A/H$ funnel
(blue shading) is important. The NUHM1 best-fit point is in a high-mass region where $\Och$ is determined by coannihilations
of nearly-degenerate $\neu1$, $\cha{1}$ and $\neu2$ \cite{coann,edsjo} (green shading),
 since $\mu \ll m_{1/2}$ and the LSP is nearly a pure higgsino.
 
 In the case of the NUHM2, all four of the mechanisms (\ref{shadings}) come into play,
 as we see in the lower right panel of Fig.~\ref{fig:m0m12}. As in the cases of the CMSSM and NUHM1,
 there are regions where stau coannihilation (pink), rapid annihilation via $A/H$ bosons (blue)
 and $\cha{1}$ coannihilations (green) are important. We also see two small bands with
 $(m_0, m_{1/2}) \sim (2000, 600) \gev$ where stop coannihilation \cite{stop} is important.
  
\begin{figure*}[htb!]
\begin{center}
\resizebox{10cm}{!}{\includegraphics{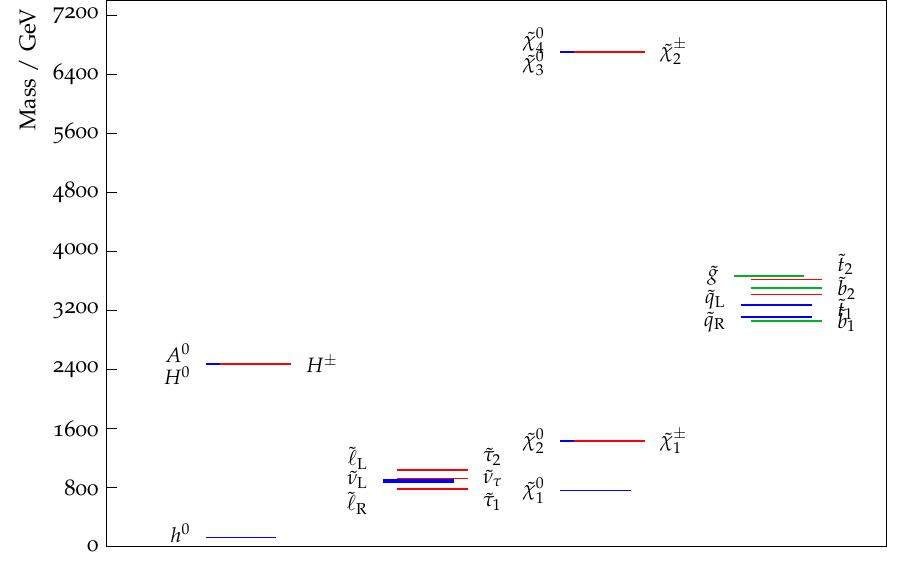}}
\end{center}
\vspace{-1cm}
\caption{\it The spectrum at the best-fit point found in our frequentist fit to the NUHM2.}
\label{fig:spectrum}
\end{figure*}

Our best-fit point for the NUHM2 has $m^2_0 < 0$ in the pink region
where the relic density is fixed by stau coannihilation. 
 As can be seen in Fig.~\ref{fig:spectrum}, the LSP and the lighter stau are indeed very nearly degenerate
 at this point, with the other sleptons only slightly heavier but the other sparticles significantly
 more massive. Also, $\MA \gg 2 \mneu{1}$, so there is no
 significant enhancement of LSP annihilations via direct-channel resonances. We emphasize,
 however, that the NUHM2 spectrum is poorly determined, and that this and other processes
 play important roles in other parts of the NUHM2 parameter space. We find $\Mh = 124.8~\gev$
 at the best-fit point.
 
\begin{figure*}[htb!]
\resizebox{8cm}{!}{\includegraphics{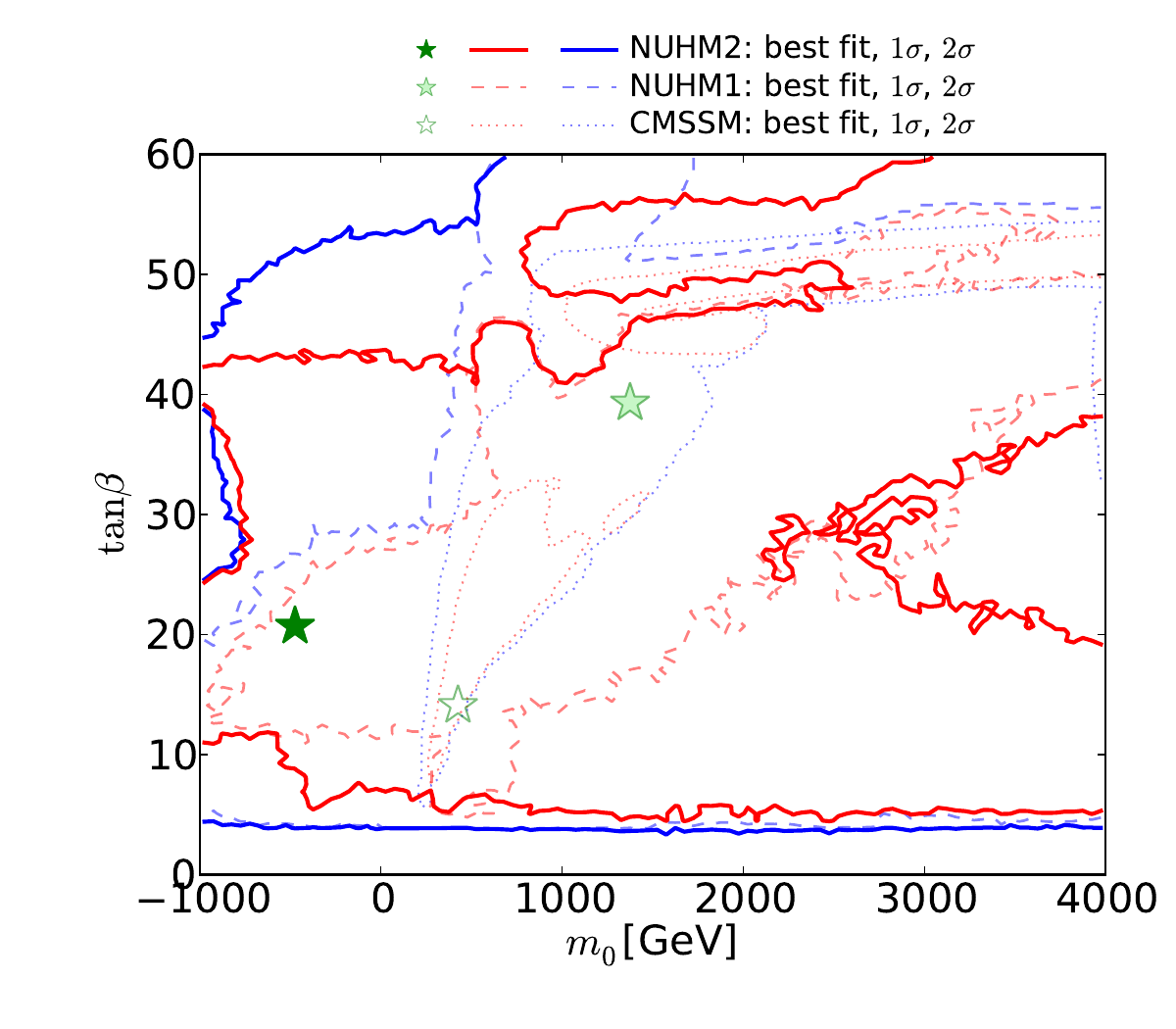}}
\resizebox{8cm}{!}{\includegraphics{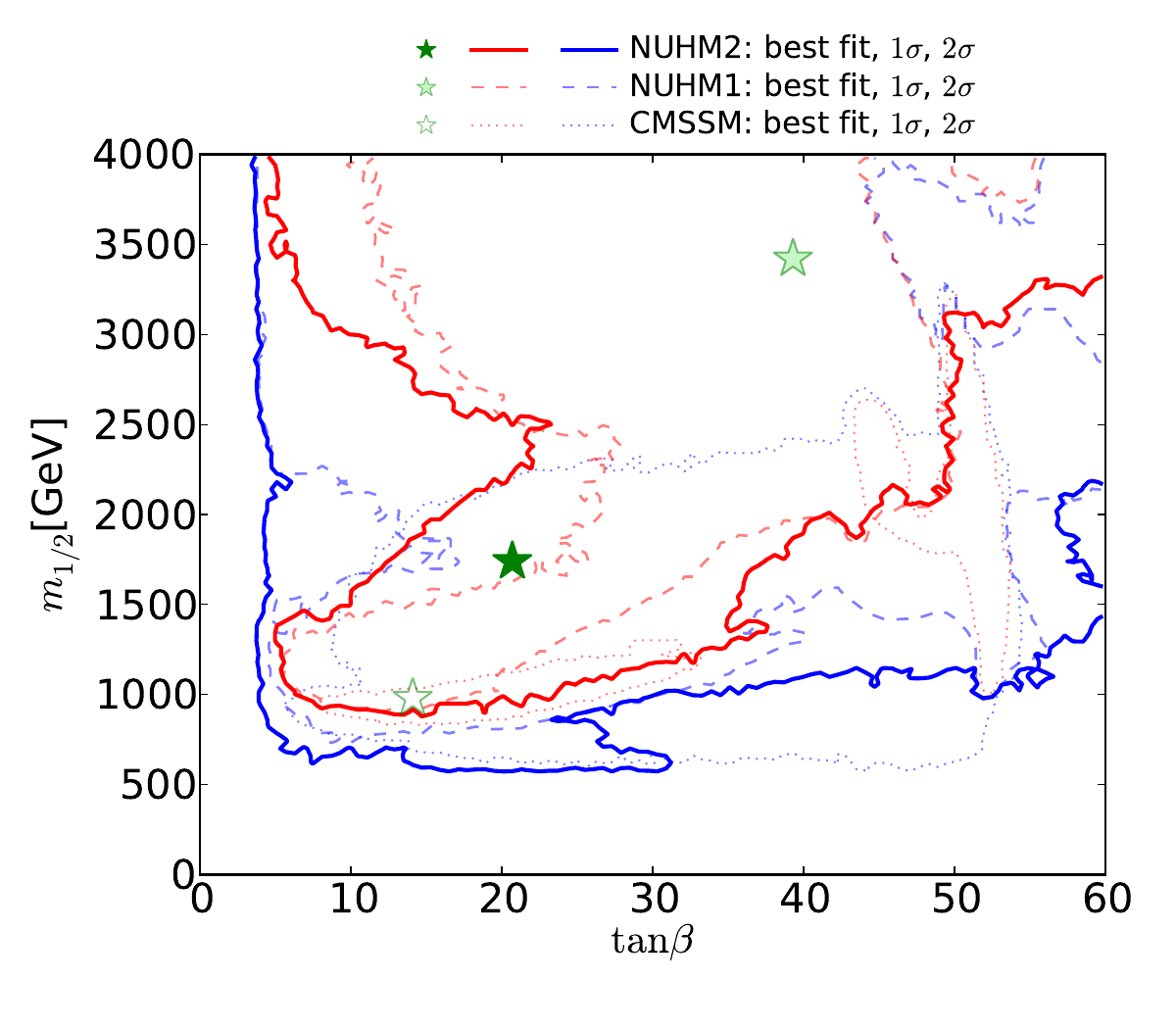}}
\caption{\it The $(\tb, m_{0})$ and $(\tb, m_{1/2})$ planes in the NUHM2, CMSSM and NUHM1. The
stars and contours have the same significations as in Fig.~\protect\ref{fig:m0m12}.}
\label{fig:m0m12tb}
\end{figure*}

~\\
\noindent
{\it Other parameter planes:}\\
Fig.~\ref{fig:m0m12tb} displays the $(m_0, \tb)$ plane (left)
and the $(\tb, m_{1/2})$ plane (right) in the NUHM2, CMSSM and NUHM1. In both panels,
we see that a large range $5 \lsim \tb \lsim 60$ is allowed at the 68\% CL (solid red contour)~\footnote{We
do not show results for $\tb > 60$ where the RGE results are less reliable.}.
The range of $\tb$ within the 68\% CL region is restricted to values $\lsim 40$ for the lower-mass lobe in 
Fig.~\ref{fig:m0m12} where $m_0 \lsim 1000 \gev$ and $m_{1/2} \lsim 2500 \gev$. Once
again, we see that the additional freedom of being able effectively to choose $\mu$ and $\MA$ independently
allows solutions with the correct relic density over a wider range of
the parameters $m_0, m_{1/2}$ and $\tan \beta$. The region of the $(m_0, \tb)$ plane with $|m_0| \lsim 1000 \gev$ is
generally in the stau coannihilation region, whereas in the region at larger $m_0$ and $\tb \lsim 40$
$\cha{1}$ coannihilation is important. The prominent horizontal
lobe in the left-hand plot at $\tan\beta \sim 50$ is associated with the
$A$-funnel region.

\begin{figure*}[htb!]
\resizebox{8cm}{!}{\includegraphics{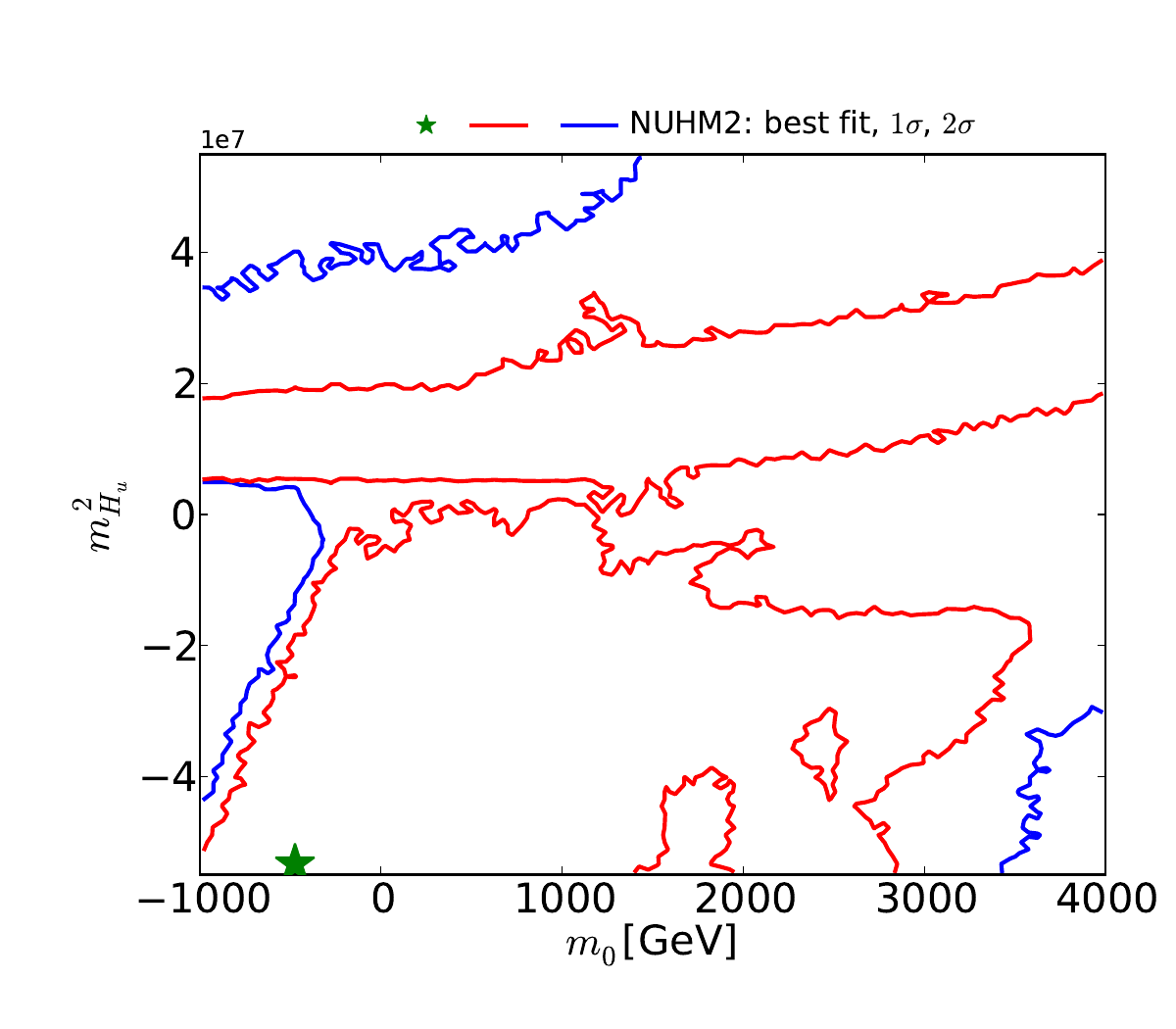}}
\resizebox{8cm}{!}{\includegraphics{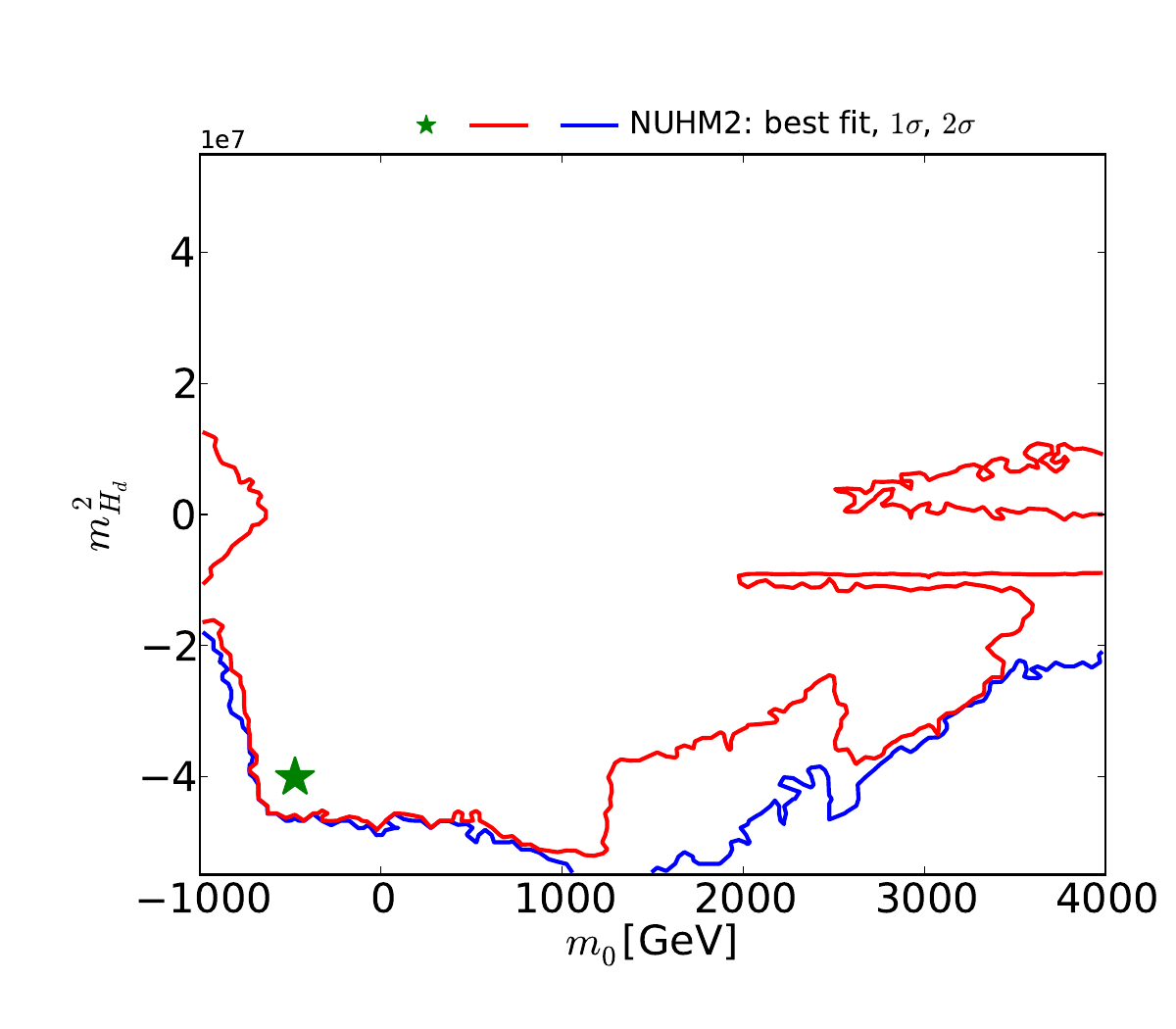}}
\caption{\it The $(m_0, m_{H_{u}}^2)$ plane (left panel) and the $(m_0,
  m_{H_{d}}^2)$ plane (right panel) in the NUHM2 fit. The significations of the solid lines and filled
  stars are the same as in Fig.~\protect\ref{fig:m0m12}.}
\label{fig:m0mHumHd}
\end{figure*}

\begin{figure*}[htb!]
\begin{center}
\resizebox{8cm}{!}{\includegraphics{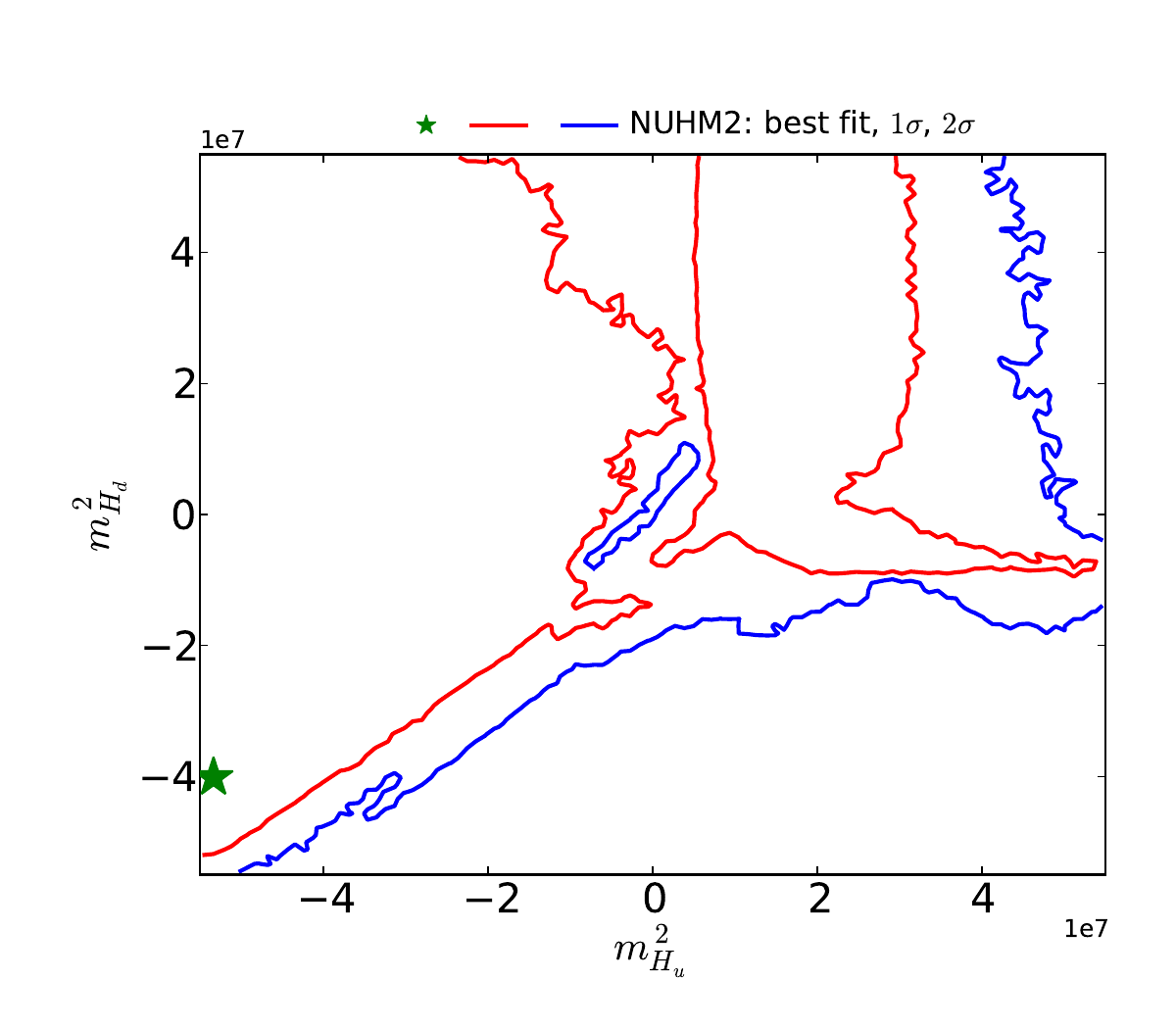}}
\end{center}
\vspace{-1cm}
\caption{\it The $(m_{H_u}^2, m_{H_d}^2)$ plane in the NUHM2. The
star and contours have the same significations as in
Fig.~\protect\ref{fig:m0m12}.}
\label{fig:mHu2mHd2}
\end{figure*}

Fig.~\ref{fig:m0mHumHd} displays the $(m_0, m_{H_u}^2)$ and $(m_0,
m_{H_d}^2)$ planes of the NUHM2 (left and right panels, respectively).
We see again that the best-fit point has $m_0 < 0$, and that both
$m_{H_{u,d}}^2 < 0$ are favoured, with a preference for $m_{H_u}^2 < m_{H_d}^2$~\footnote{However,
the exact locations of the CL contours near the best-fit point in the right panel of Fig.~\ref{fig:m0mHumHd} are subject
to our sampling restrictions.}. The reason for this preference can
be understood from (\ref{defS}). To obtain a neutralino LSP, we require $S < 0$, which then requires 
$m_{H_u}^2 < m_{H_d}^2$. In general, stau coannihilation is most important when  $m_{H_u}^2$ or $m_{H_d}^2 \lsim 0$,
whereas $\cha{1}$ coannihilation is more important when $m_{H_u}^2$ or $m_{H_d}^2 \gsim 0$.
Fig.~\ref{fig:mHu2mHd2} displays the $(m_{H_u}^2, m_{H_d}^2)$ plane for the NUHM2, where we
see that the best-fit point has $m_{H_u}^2 < m_{H_d}^2 < 0$. However, we emphasize that
the global likelihood function is quite flat in $m_{H_{u,d}}^2$, and the
most reliable statement that can be made is that the quadrant $m_{H_u}^2 > 0, m_{H_d}^2 < 0$
is the least favoured. When $m_{H_u}^2 \lsim 0$, stau coannihilation is important  for $m_{H_d}^2 \gsim m_{H_u}^2$,
but the $A/H$ funnel is important when $m_{H_d}^2 \sim m_{H_u}^2$. When $m_{H_u}^2 \gsim 0$,
$\cha{1}$ coannihilation is important for $m_{H_d}^2 \gsim 0$ whereas stop coannihilation becomes
important for $m_{H_d}^2 < 0$.

\begin{figure*}[htb!]
\resizebox{8cm}{!}{\includegraphics{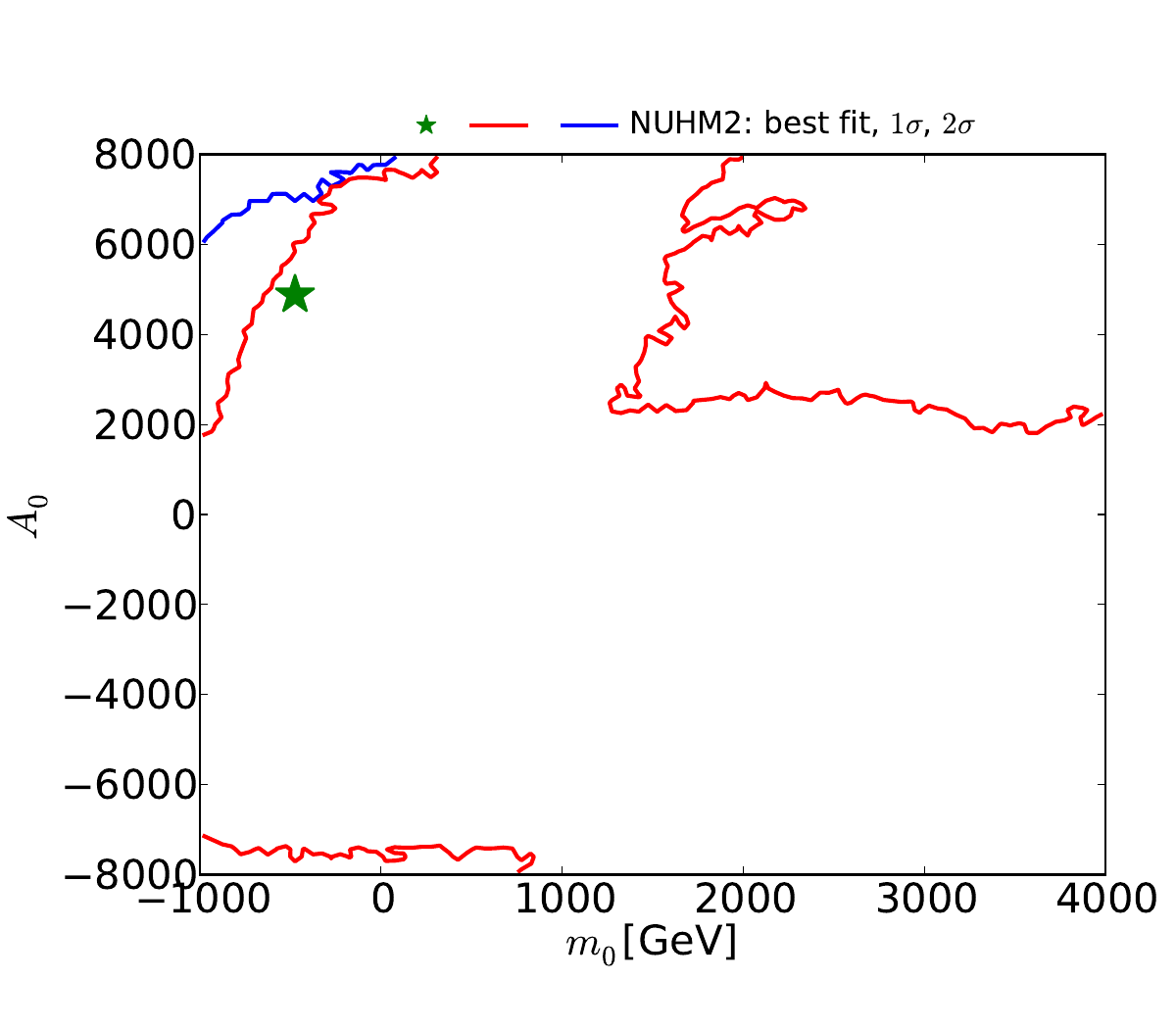}}
\resizebox{8cm}{!}{\includegraphics{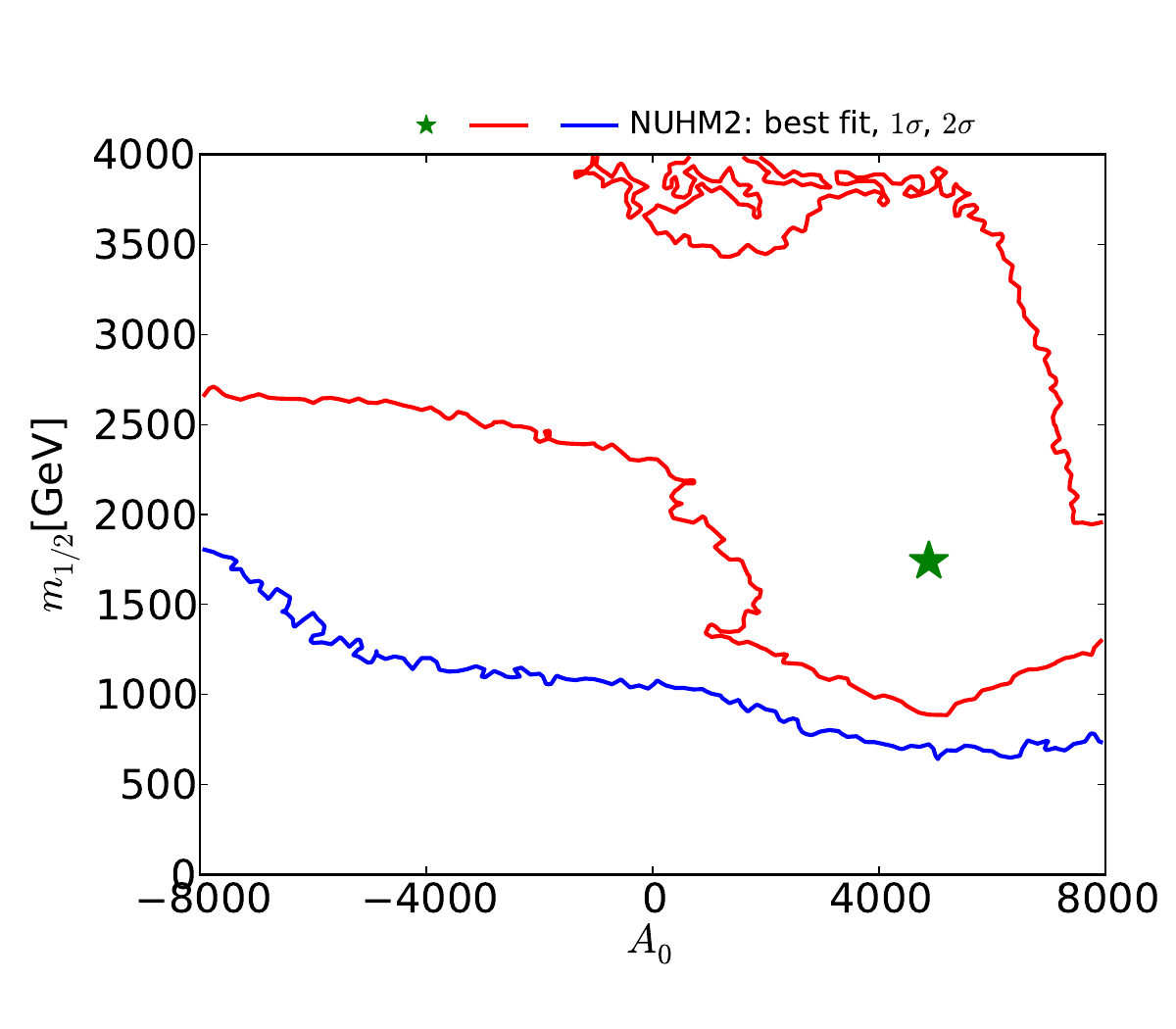}}
\caption{\it The $(m_{0}, A_0)$ plane (left panel) and the $(m_{1/2}, A_0)$ plane (right panel) in the NUHM2.
The significations of the solid lines and filled stars are the same as in Fig.~\protect\ref{fig:m0m12}.}
\label{fig:m0m12A0_ko}
\vspace{1em}
\end{figure*}

Fig.~\ref{fig:m0m12A0_ko} displays the $(m_{0}, A_0)$ plane (left)
and the $(A_0, m_{1/2})$ plane (right) for the NUHM2.
The fit does not exhibit any overall preference for a sign of $A_0$.
However, we see that negative values of $A_0$ are generally preferred when $m_0$ and $m_{1/2}$ are large, 
whereas the low-mass lobe in
Fig.~\ref{fig:m0m12} is generally associated with positive values of $A_0$\footnote{We recall that we use the same convention for
the sign of $A_0$ as in~\cite{mc8,mc9}, which is opposite to the
convention used in, e.g., {\tt SoftSUSY}.}.
This tendency is driven by the value of $\Mh$ measured at the LHC.

\begin{figure*}[htb!]
\begin{center}
\resizebox{8cm}{!}{\includegraphics{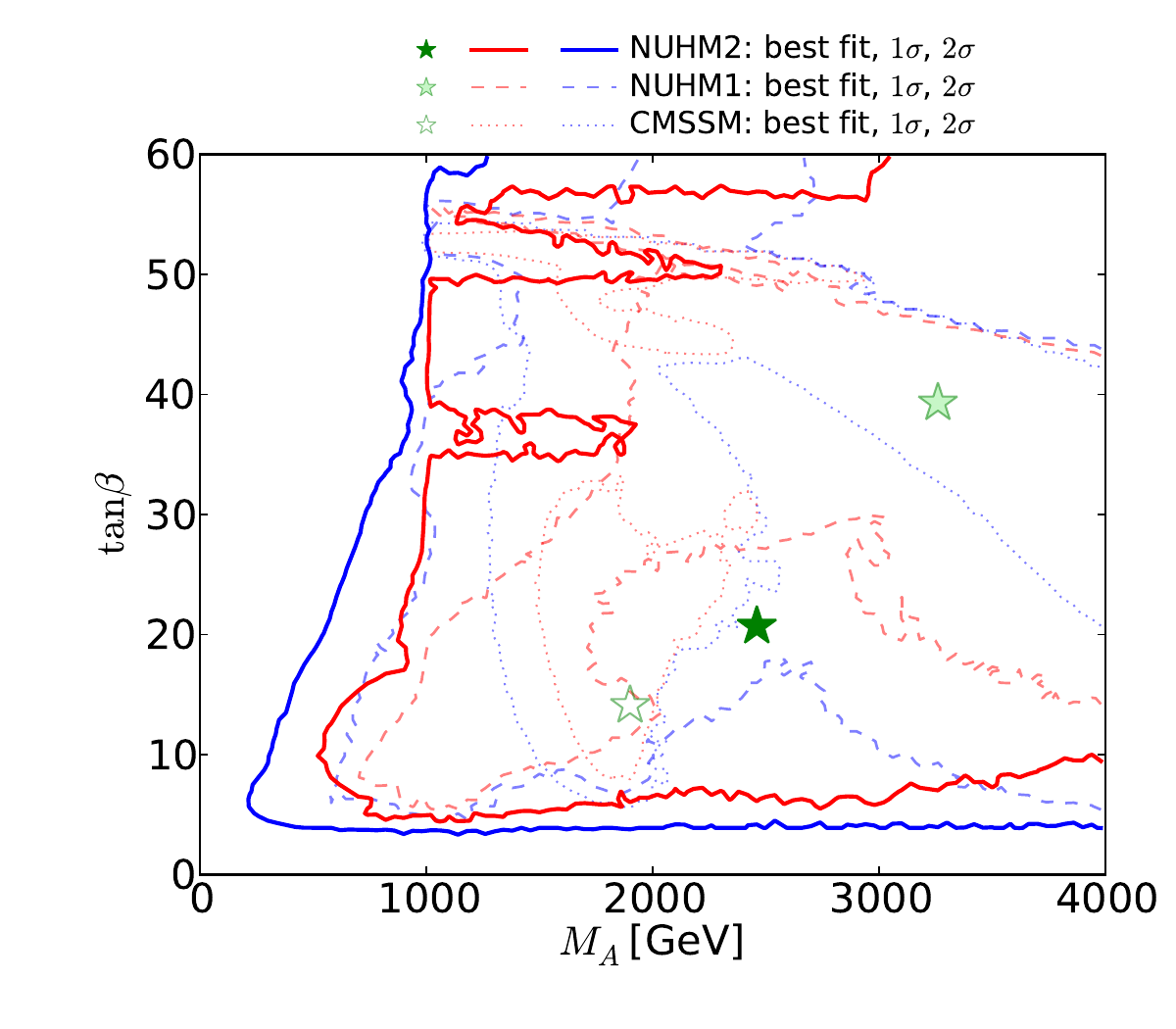}}
\end{center}
\vspace{-1cm}
\caption{\it The $(\MA, \tan \beta)$ plane in the NUHM2, CMSSM and NUHM1. The lines and
  stars have the same significations as in Fig.~\protect\ref{fig:m0m12}.
}
\label{fig:MAtb}
\end{figure*}

Fig.~\ref{fig:MAtb} displays the $(\MA, \tan \beta)$ plane in the NUHM2 (solid lines),
CMSSM (dashed lines) and NUHM1 (dotted lines). In the NUHM2 we see a 95\% CL lower limit
on $\MA$ that increases from $\sim 200 \gev$ when $\tb \sim 5$ to $1000 \gev$ when $\tb \sim 50$,
which is essentially determined by the $H/A \rightarrow \tau \tau$ constraint~\cite{ATLASHA}.
The best-fit value of $\MA \sim 2500 \gev$,
but the global $\chi^2$ function is very flat, and this model parameter is not well determined,
and could be as low as 500~GeV at the 68\% CL.
We find a 95\% CL lower limit $\tb \gsim 4$, which is quite insensitive to the value of $\MA$.
We find that $\cha{1}$ coannihilation is generally important for $\MA \lsim 2000 \gev$,
whereas stau coannihilation is important for $\MA \gsim 2000 \gev$. The $A/H$ funnel becomes
important for $\MA \sim 2000 \gev$, and also for $\tb \gsim 50$.

\subsection{Summary of NUHM2 Global Fit}

Table~\ref{tab:bestfits} summarizes our results for the our best-fit points
in a global fit to the NUHM2,
compared with fits in the NUHM1 and the CMSSM using the same post-LHC Run~1 data set.
We see that the total $\chi^2$ in the best NUHM2 fit is lowered by only $\Delta \chi^2=0.2$ 
from the best NUHM1 fit, so the extra parameter in the NUHM2 does not provide 
a significant advantage. According to the F-test,
there is a 77\% chance that the data are represented better by the NUHM1 than by the CMSSM,
whereas there is only a 28\% chance that the NUHM2 is an improvement on the NUHM1,
and a 78\% chance that the NUHM2 represents the data better than the CMSSM. None of these
can be regarded as significant.

We note that the NUHM2 best-fit value of $m_0$ is small and negative,
and that it is accompanied by values of $m_{H_u}^2$ and
$m_{H_d}^2$ that are also negative and larger in magnitude. The best-fit value
of $m_{1/2}$ in the NUHM2 lies significantly beyond the direct lower limit from
sparticle searches at the LHC.
We also find that a positive value of $A_0$ is preferred, in contrast to
the NUHM1 and the CMSSM which have much larger values of $m_0$ and $m_{1/2}$ at their 
best fit points.
That said, we point out that the likelihood functions are extremely shallow, and the 68\% ranges very large,
so the best fit point should not be over-interpreted.

\begin{table*}[!tbh!]
\renewcommand{\arraystretch}{1.5}
\begin{center}
\begin{tabular}{|c||c|c|c|c|c|c|c|c|c|} \hline
Model & $\chi^2$/d.o.f. & Prob- & $m_0$ & $m_{1/2}$ & $A_0$ & $\tb$ & $m_{H_u}^2$ & $m_{H_d}^2$ \\
  &    & ability & (GeV) & (GeV) & (GeV) & & ($\gev^2$) & ($\gev^2$) \\ 
\hline \hline
CMSSM & 35.0/23 & 5.2\% & 420 & 970 & 3000 & 14 & $= m_0^2$ & $= m_0^2$ \\
\hline
NUHM1 & 32.7/22 & 6.6\% &1380 &3420 & -3140 & 39 & $1.33 \times 10^7$ & $= m_{H_u}^2$ \\
\hline 
NUHM2 & 32.5/21 & 5.2\% & -490 & 1730 & 4930 & 21 & $- 5.28 \times 10^7$ & $- 4.03 \times 10^7$\\
\hline
\end{tabular}
\caption{\it The best-fit points found in global fits in the CMSSM, the NUHM1 and the NUHM2,
  using the same experimental constraints (and their theoretical interpretations):
   the difference in the CMSSM best-fit from that found in~\cite{mc9}
is due to using the updated ATLAS jets + $\ETslash$ constraint~\cite{ATLAS20}. We note that the
  overall likelihood functions in all the models are quite flat, so that the precise
  locations of the best-fit points are not very significant, and for
  this reason we do not quote uncertainties.} 
\label{tab:bestfits}
\end{center}
\end{table*}


\section{Predictions for Physical Observables}

We now turn to the predictions for physical observables
that emerge from our frequentist analysis of the NUHM2
parameter space, and compare them with corresponding
predictions from our previous analyses of the CMSSM and
NUHM1 parameter spaces~\cite{mc9}. Since the CMSSM is a subset of the NUHM1, which is
itself a subset of the NUHM2, $\chi^2|_{\rm CMSSM} \ge \chi^2|_{\rm NUHM1} \ge \chi^2|_{\rm NUHM2}$
everywhere. However, this is not immediately visible in the plots below,
in which we plot the difference $\Delta \chi^2$ from the minimum value of
$\chi^2$ in that model shown in the Table.

\subsection{Sparticle Masses}

\begin{figure*}[htb!]
\resizebox{8cm}{!}{\includegraphics{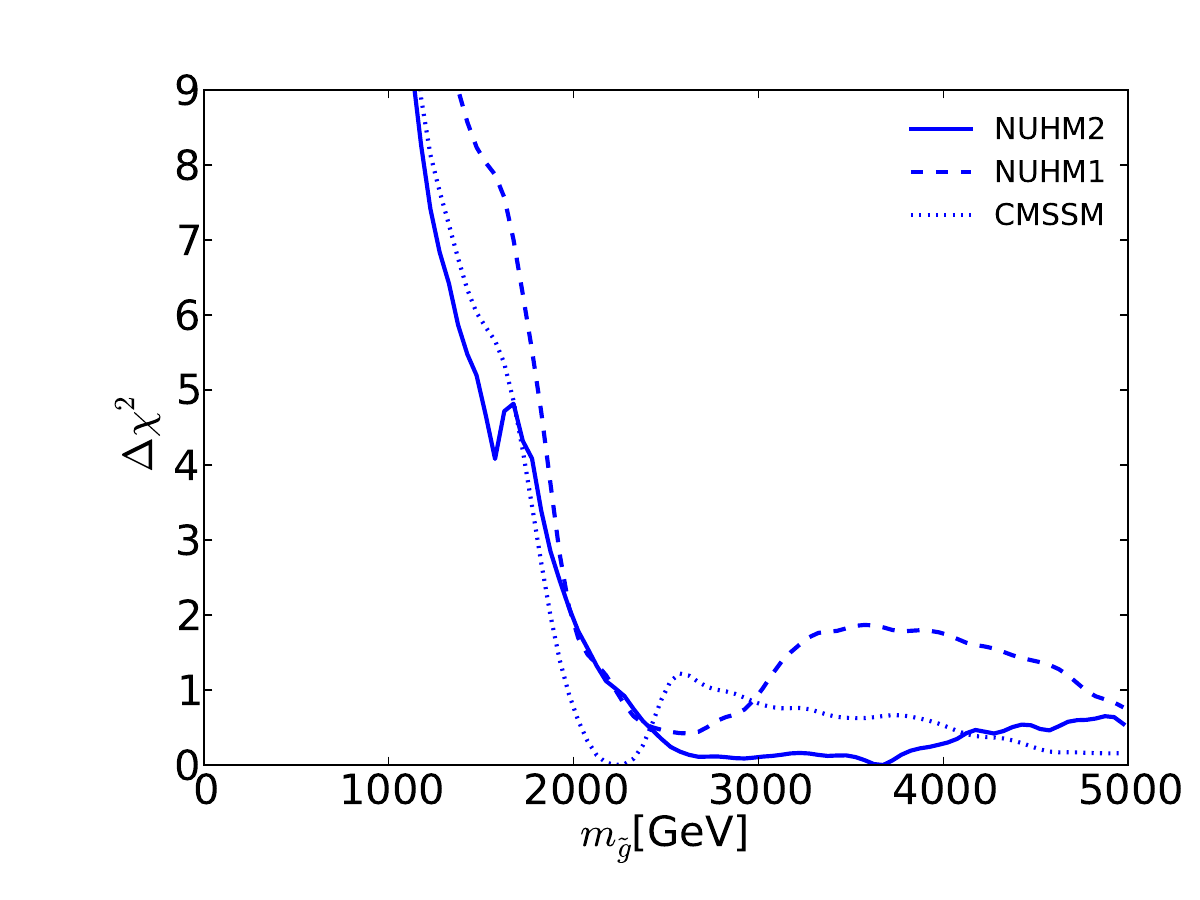}}
\resizebox{8cm}{!}{\includegraphics{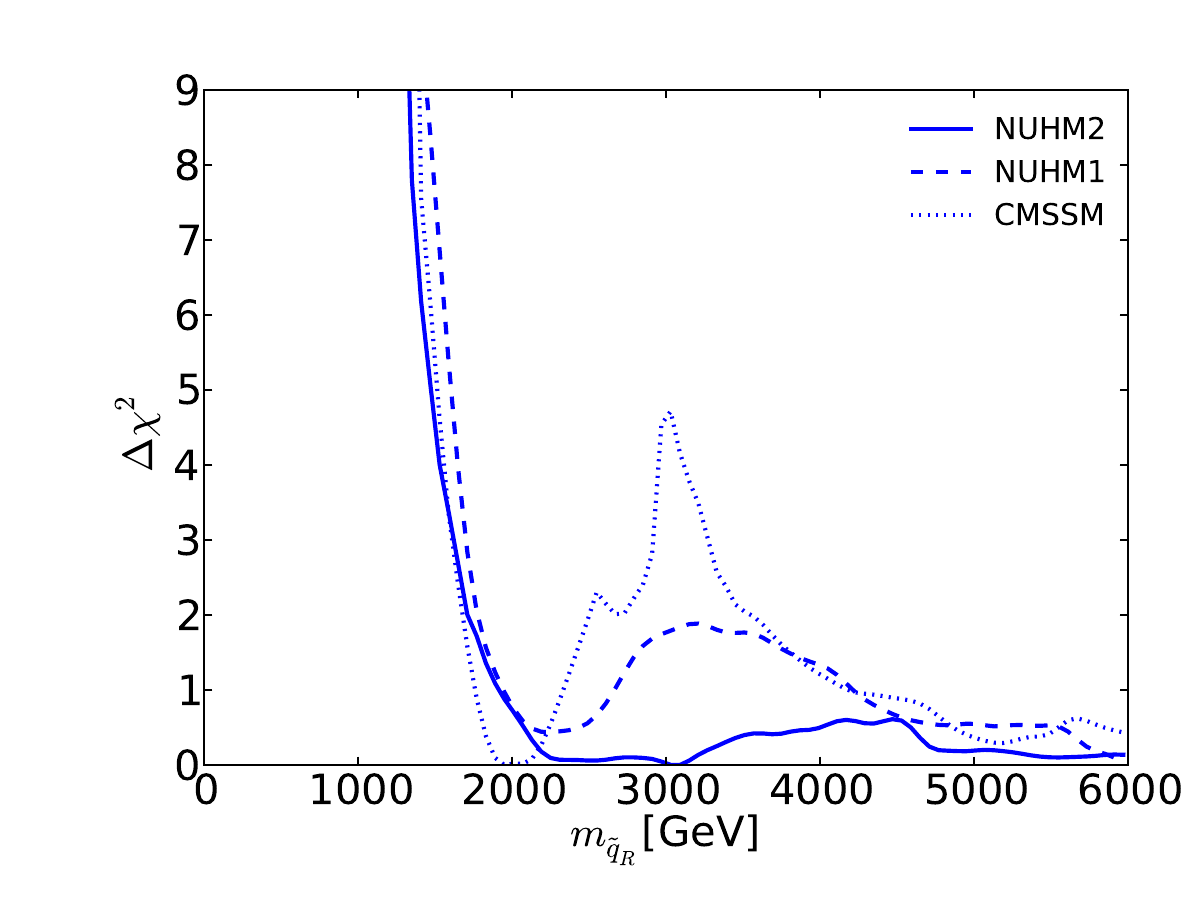}} \\
\vspace{-1cm}
\caption{\it The $\Delta \chi^2$ likelihood function in the NUHM2 (solid line) as a function of
  $\mgl$ (left panel) and $\msqr$ (right panel). The dotted (dashed) lines are
  for the corresponding fits in the CMSSM and NUHM1, respectively.
}
\label{fig:mglmsq}
\end{figure*}

In the left panel of Fig.~\ref{fig:mglmsq} we display the $\Delta \chi^2$ 
function in the NUHM2 (solid line) as a function of $\mgl$. We see that $\mgl \gsim 1.5 \tev$
is preferred, as was the case in the CMSSM and NUHM1, at the 95\% CL, 
and that the $\Delta \chi^2$ function is quite flat for $\mgl \gsim 2.5 \tev$.
The lower limit on $\mgl$ is mainly due to the ATLAS jets + $\ETslash$ constraint,
counteracted to some extent by \gmt: the LHC $\Mh$ measurement plays no role.
The best-fit point has $\mgl \sim 3670~\gev$
as seen also in Fig.~\ref{fig:spectrum}.
At low masses, the $\Delta \chi^2$ function is similar to that for the CMSSM (dotted line),
and also to the NUHM1(dashed line) when $\mgl \lsim 2 \tev$. Above this mass, the difference between the $\Delta \chi^2$ functions
for the NUHM2 and the NUHM1 is largest for $3 \tev \lsim \mgl \lsim 5 \tev$. 

The right panel of Fig.~\ref{fig:mglmsq} displays the $\Delta \chi^2$ likelihood as a function 
of $\msqr$, defined here to be the average of the spartners of the right-handed
components of the four lightest quarks. We see that $\msq \gsim 1.5 \tev$ at the
95\% CL in the NUHM2, driven essentially by the ATLAS jets + $\ETslash$
constraint, with a best-fit value $\msqr \sim 3080~\gev$ as seen also in Fig.~\ref{fig:spectrum},
and that the $\Delta \chi^2$ function in this model is very similar
to those in the NUHM1 and CMSSM for $\msqr \lsim 2 \tev$. However, the $\Delta \chi^2$
functions in these models differ quite significantly for $2 \tev \lsim \msqr \lsim 4.5 \tev$,
reflecting the fact visible in Fig.~\ref{fig:m0m12} that the separation between the low- and
high-mass regions becomes less pronounced as the Higgs mass universality is
progressively relaxed. This can be traced back to the broader range of options for bringing
the cold dark matter density into the range preferred by cosmology. 

\begin{figure*}[htb!]
\resizebox{8cm}{!}{\includegraphics{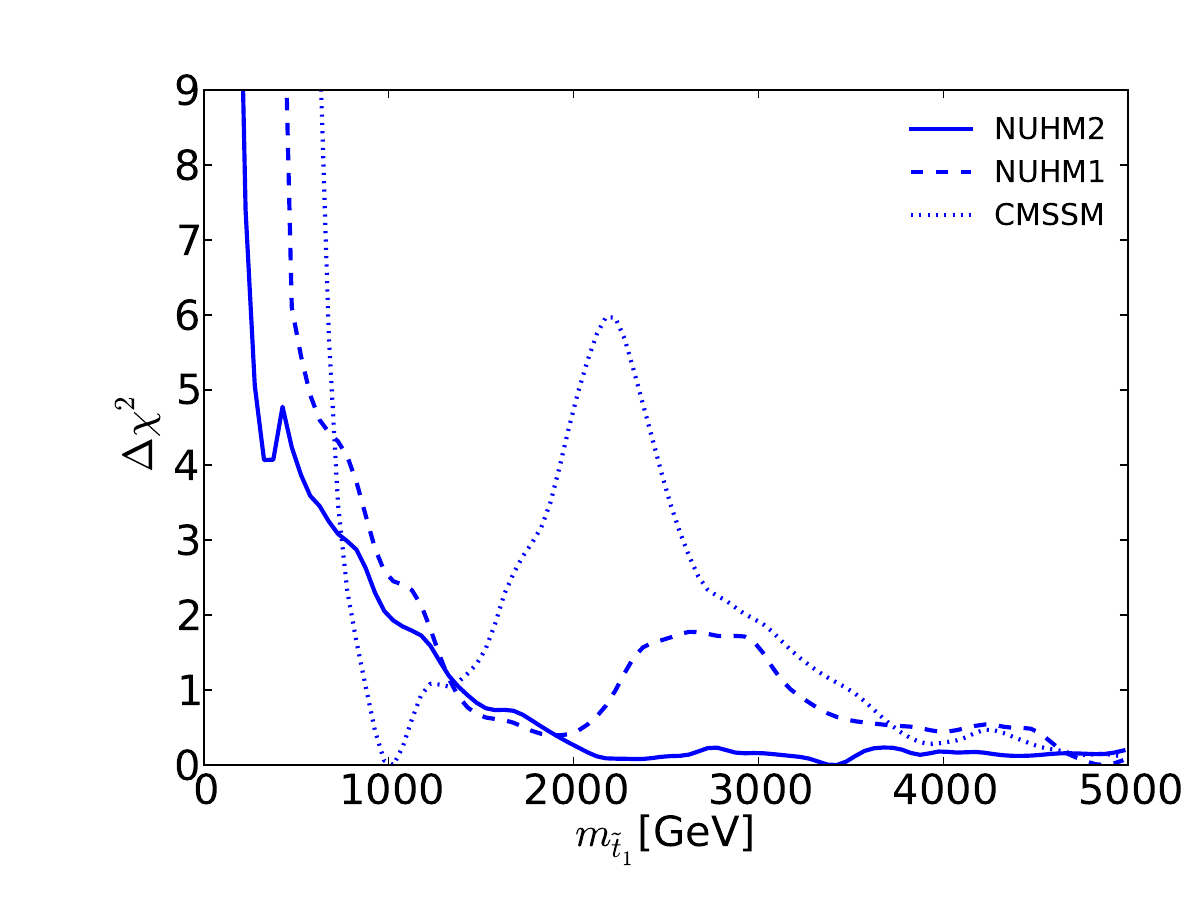}}
\resizebox{8cm}{!}{\includegraphics{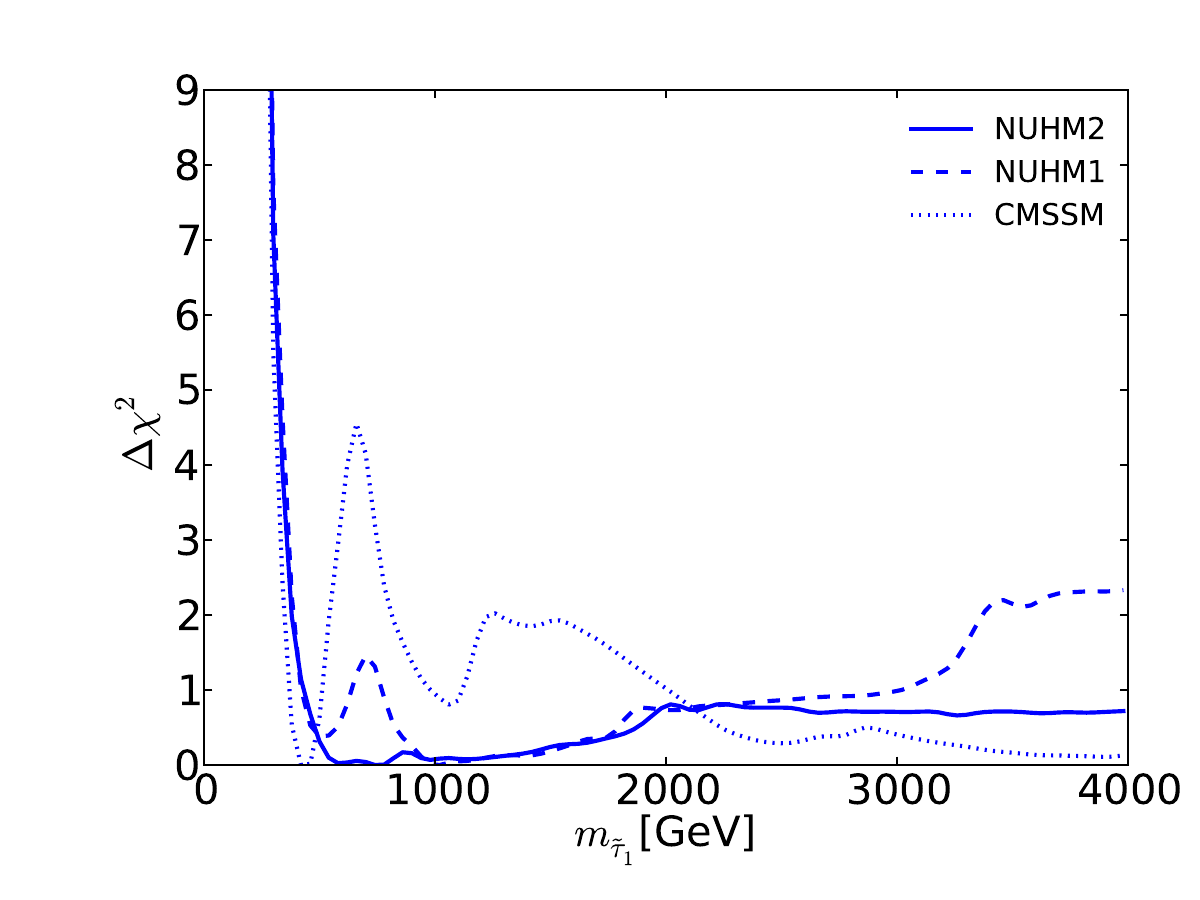}}
\vspace{-1cm}
\caption{\it As in Fig.~\protect\ref{fig:mglmsq}, for $m_{\tilde t_1}$ (left panel)
and for $m_{\tilde \tau_1}$ (right panel).
} 
\label{fig:mstopmstau}
\end{figure*}

In the left panel of Fig.~\ref{fig:mstopmstau} we display the $\Delta \chi^2$
likelihood as a function of $m_{\tilde t_1}$.
In this case the lower mass limit is not driven by the ATLAS jets + $\ETslash$ search.
On the other hand, the $\Delta \chi^2$ functions for these models are quite different at both
larger and smaller $m_{\tilde t_1}$: lower masses are not so strongly disfavoured in the NUHM2,
and the features found in the NUHM1 and CMSSM at $m_{\tilde t_1} \sim 1 \tev$ and $\in (2, 3) \tev$
are not found in the NUHM2, whose $\Delta \chi^2$ function falls almost monotonically as $m_{\tilde t_1}$
increases. This reflects again the fact that the low- and
high-mass regions are less distinct in the NUHM2.
There are also some stop coannihilation points at low $m_{\tilde t_1}$.
The best-fit point has $m_{\tilde t_1} \sim 3420~\gev$ as seen also in Fig.~\ref{fig:spectrum}.
The right panel of Fig.~\ref{fig:mstopmstau} displays the $\Delta \chi^2$ likelihoods
in the NUHM2, NUHM1 and CMSSM  as functions of $m_{\tilde \tau_1}$. At low mass, 
we see that the $\Delta \chi^2$ functions are almost identical in the three models, giving a lower
bound $m_{\tilde \tau_1} \gsim 300 \gev$ at the 95\% CL, driven by the ATLAS
jets + $\ETslash$ search. 
At higher masses, the structures seen in
the $\Delta \chi^2$ functions for the NUHM1 (dashed line) and CMSSM (dotted line)
are absent for the NUHM2, whose $\Delta \chi^2$ function (solid line)
has a shallow minimum at $m_{\tilde \tau_1} \sim 780 \gev$. 

\begin{figure*}[htb!]
\resizebox{8cm}{!}{\includegraphics{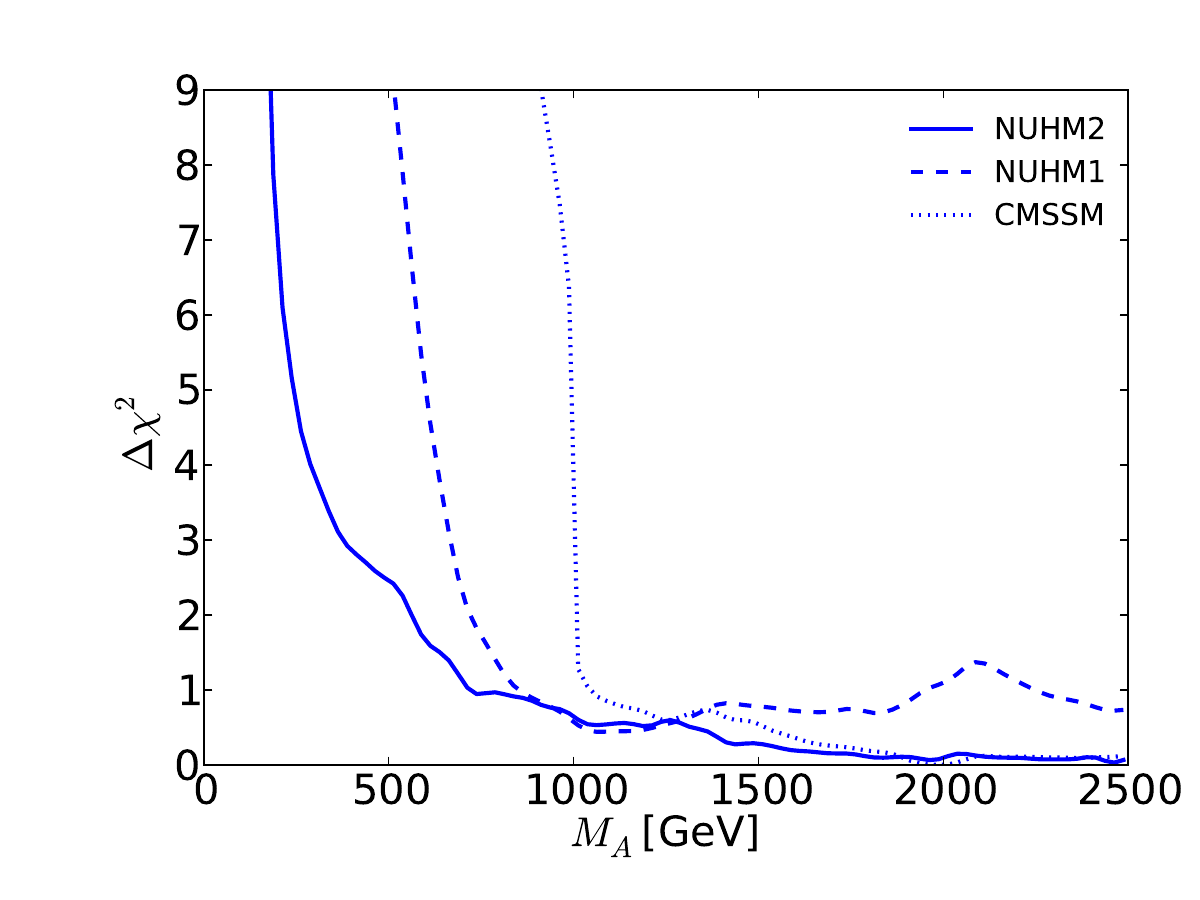}}
\resizebox{8cm}{!}{\includegraphics{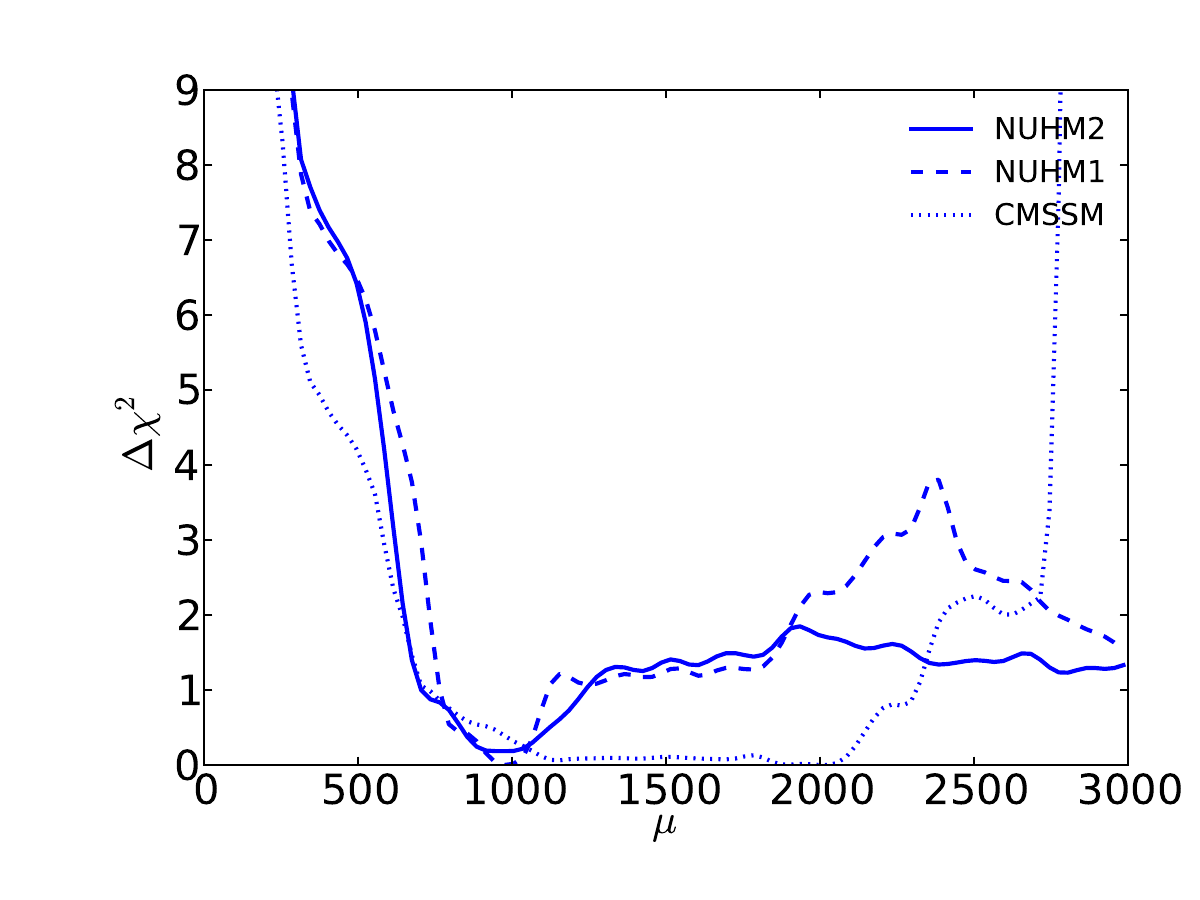}}
\vspace{-1cm}
\caption{\it As in Fig.~\protect\ref{fig:mglmsq}, for $\MA$ (left panel)
and for $\mu$ (right panel).}
\label{fig:MAmu}
\end{figure*}

The left panel of Fig.~\ref{fig:MAmu} displays the dependences of the $\Delta \chi^2$
functions in the NUHM2, NUHM1 and CMSSM on $\MA$. We see that the $\Delta \chi^2$
function for the NUHM2 is quite flat above $\sim 500 \gev$, following a steep rise at
lower masses and a 95\% CL lower limit $\MA \gsim 200 \gev$.
The best-fit point has $\MA \sim 2470~\gev$ as seen also in Fig.~\ref{fig:spectrum}.
The right panel of Fig.~\ref{fig:MAmu} displays the corresponding $\Delta \chi^2$
function for $\mu$. Like $\MA$, this extra degree of freedom in the NUHM2 is
poorly constrained by current data.

\begin{figure*}[htb!]
\resizebox{8cm}{!}{\includegraphics{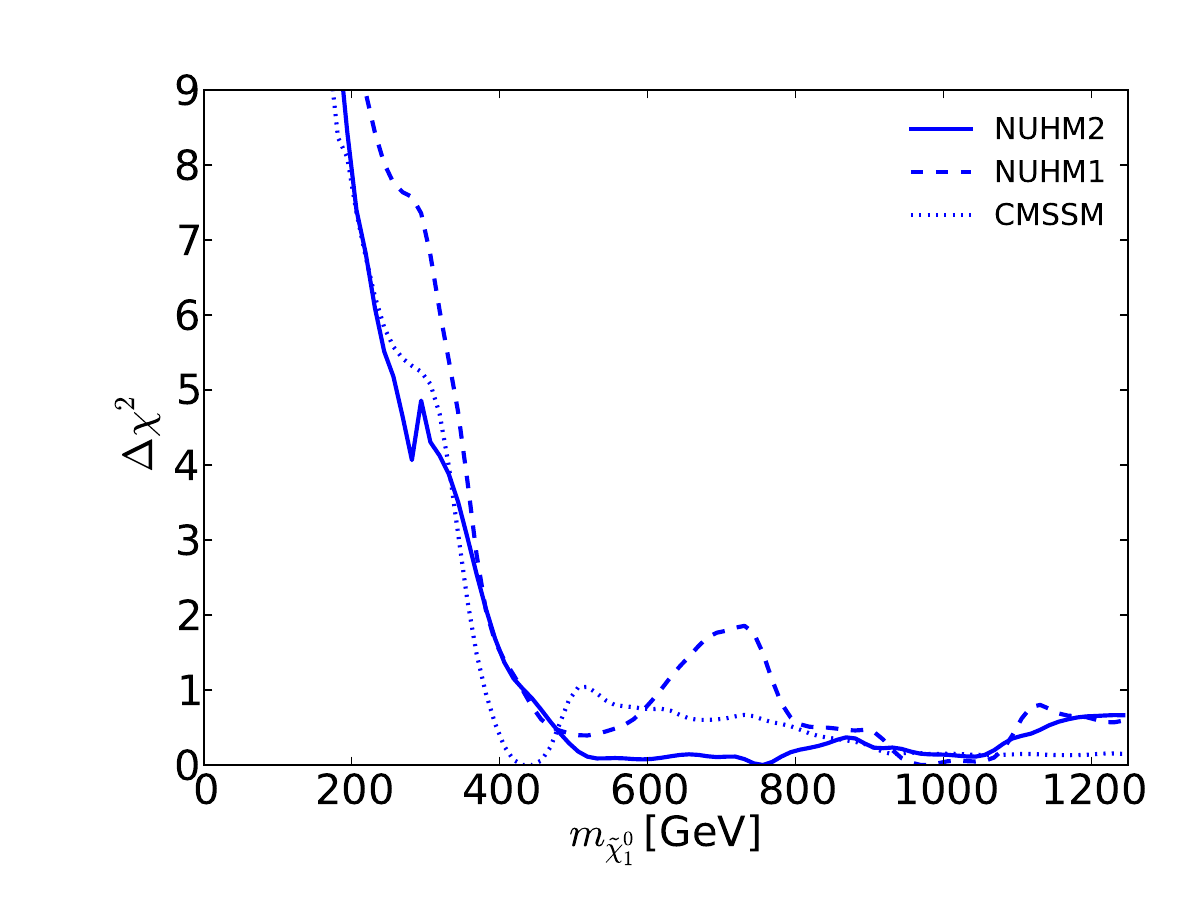}}
\resizebox{8cm}{!}{\includegraphics{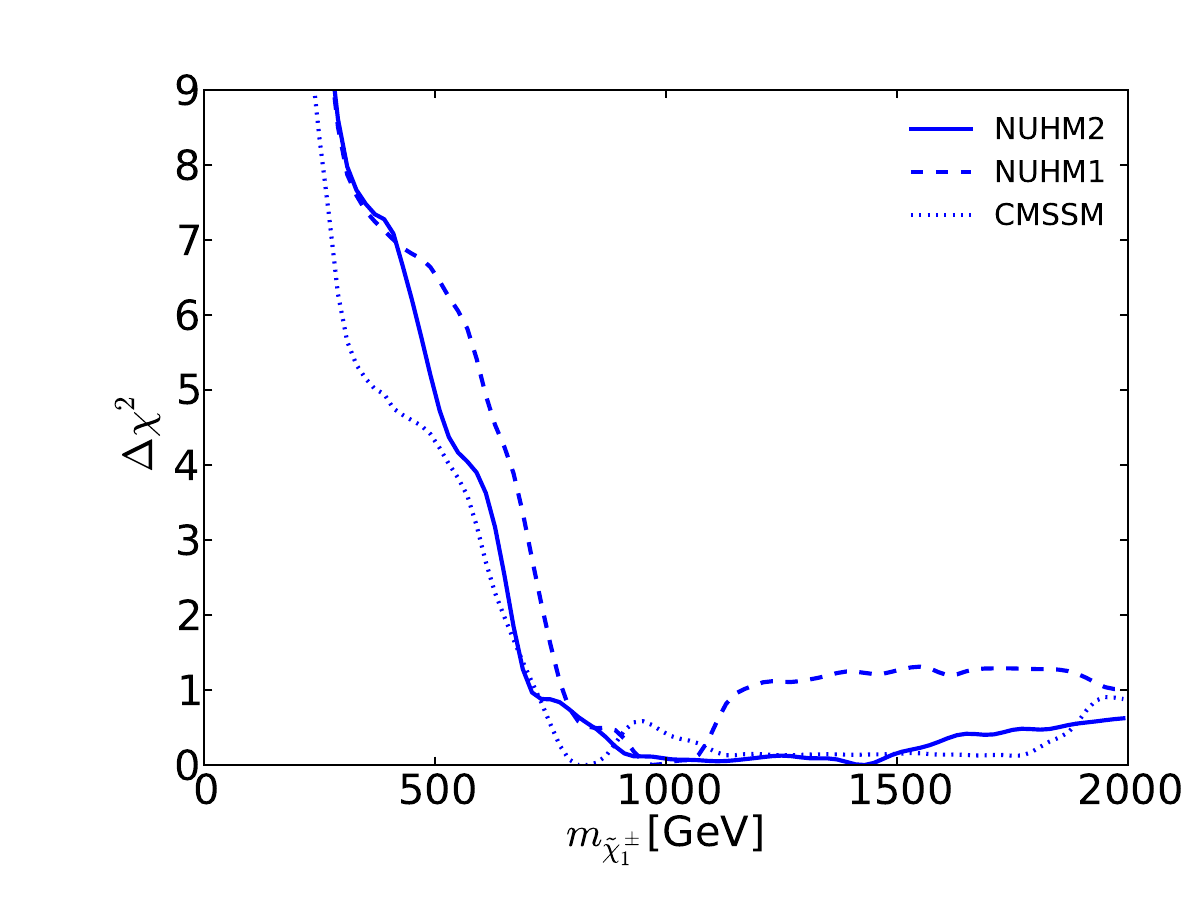}}
\vspace{-1cm}
\caption{\it As in Fig.~\protect\ref{fig:mglmsq}, for $\mneu{1}$ (left panel)
and for $\mcha{1}$ (right panel).} 
\label{fig:mchi0mchipm}
\end{figure*}

Fig.~\ref{fig:mchi0mchipm} displays the $\Delta \chi^2$ functions for
$\mneu{1}$ (in the left panel) and $\mcha{1}$ (in the right panel) in
the NUHM2, the NUHM1 and the CMSSM. The $\Delta \chi^2$ functions for
$\mneu{1}$ are quite similar at low masses, being largely driven by the ATLAS jets + $\ETslash$
constraint, and we find that
$\mneu{1} \gsim 250 \gev$ at the 95\% CL. The $\Delta \chi^2$ function in the
NUHM2 (solid line) then has a shallow minimum for $\mneu{1} \in (600,
1000) \gev$, with a best-fit value $\sim 760~\gev$. 
As already mentioned, the NUHM2 best-fit point is in the stau coannihilation region,
with $m_{\tilde \tau_1} - \mneu{1} \sim 18~\gev$ and the other sleptons slightly
heavier, as also seen in Fig.~\ref{fig:spectrum}. In the case of $\mcha{1}$, the NUHM2 $\Delta \chi^2$ function
has a 95\% CL lower bound $\gsim 500 \gev$ and  a shallow minimum for
$\mcha{1} \in (1000, 1500) \gev$ and a best-fit value $\sim 1430~\gev$ as also seen in Fig.~\ref{fig:spectrum}.
The extra degree of freedom in the NUHM2 compared to the NUHM1 does not
relax significantly the lower bounds on the $\cha{1}$ and $\neu{1}$ masses.

\begin{figure*}[htb!]
\resizebox{8cm}{!}{\includegraphics{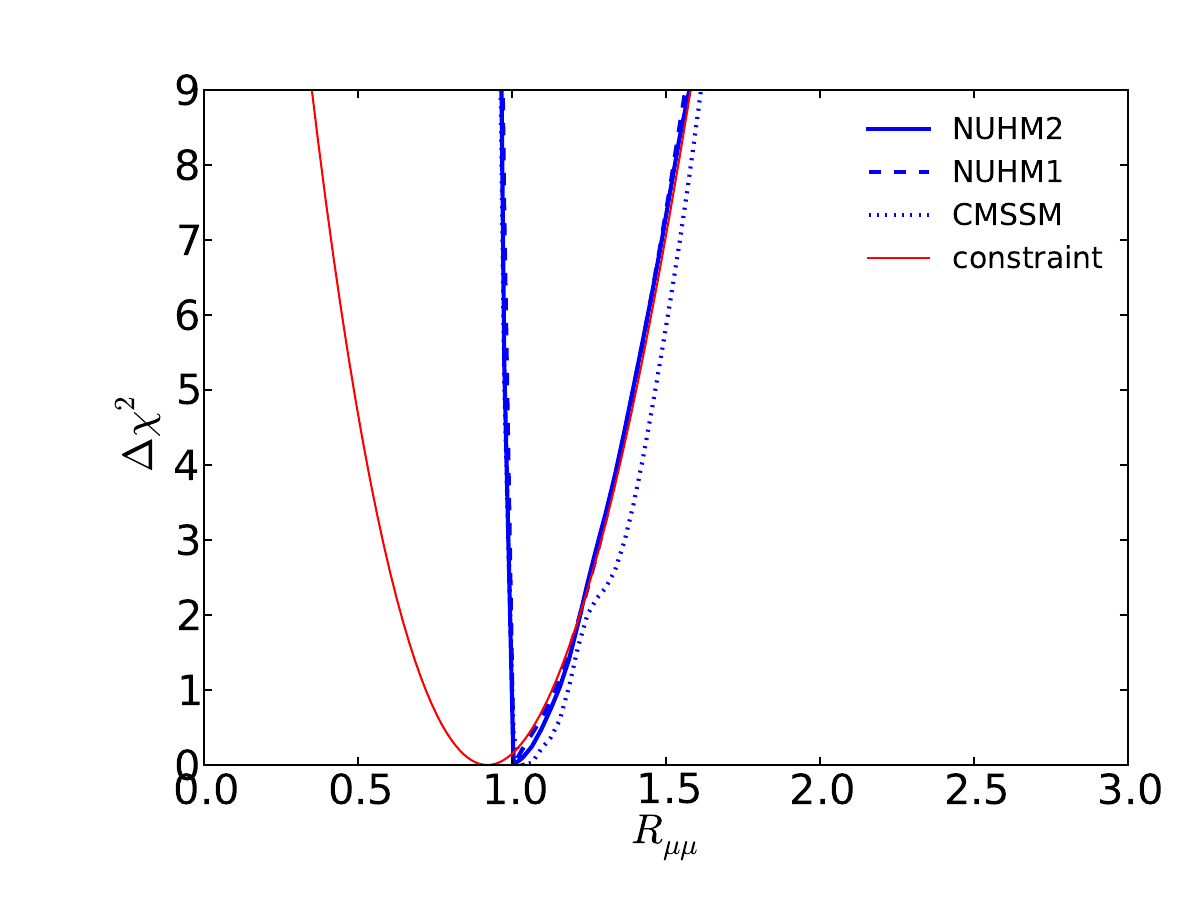}}
\resizebox{8cm}{!}{\includegraphics{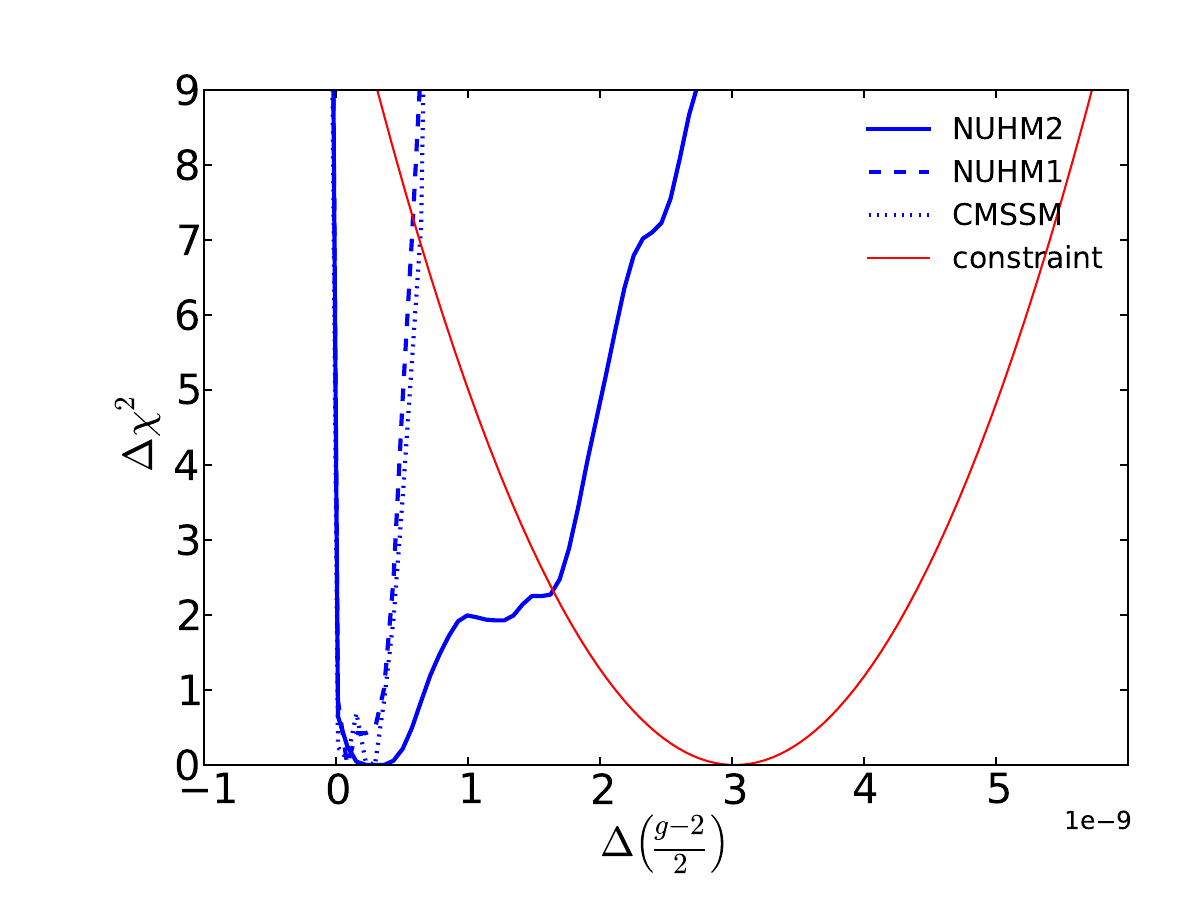}}
\vspace{-1cm}
\caption{\it As in Fig.~\protect\ref{fig:mglmsq}, for \rmm\  (left panel)
and for $\Delta \left( \frac{g-2}{2} \right)$ (right panel). In each panel, we display separately as a red line the contribution
of that individual observable to the global $\chi^2$ functions.}
\label{fig:bmmgmt}
\end{figure*}

The left panel of Fig.~\ref{fig:bmmgmt} displays the $\Delta \chi^2$ functions
for \rmm\ (defined here as $BR(B_s \to \mu^+\mu^-)/BR(B_s \to \mu^+\mu^-)_{SM}$) in the NUHM2, NUHM1 and CMSSM. We see that they are almost
identical, and that all three models allow no scope for \rmm\
to fall significantly below the SM value within the 95\% confidence level range.
For  \rmm\  above the Standard Model value, the $\Delta \chi^2$ functions
all rise in the same way as the contribution from the experimental constraint on \rmm\ (red line),
implying that the other constraints do not impose significant constraints on \rmm\ above the Standard Model value.
The fact that the CMSSM appears to have slightly larger freedom for
 \rmm\ is related to the fact the total $\chi^2$ is larger than in the
  other models. Shifting the CMSSM curve in the right panel of
  \reffi{fig:bmmgmt} to account for that difference, the CMSSM region
  would be fully contained in the NUHM1,2 regions, as expected because of the
  stronger restrictions in the CMSSM.

\subsection{The Anomalous Magnetic Moment of the Muon}

The right panel of Fig.~\ref{fig:bmmgmt} displays the $\Delta \chi^2$ functions for
the difference from the SM: $\Delta \left( \frac{g-2}{2} \right)$ in the NUHM2, NUHM1 and CMSSM, as blue solid, dashed and dotted lines, respectively.
Also shown, as a solid red line, is the \gmt\ contribution to the $\chi^2$ function. As is well known,
the other constraints, principally those from the LHC, do not allow a
large SUSY contribution to \gmt\ 
within the NUHM1 (dashed line) or the CMSSM (dotted line). 
We find that in the NUHM2 the most important role
is played by the LHC $\Mh$ measurement.
As we also see in the right panel of Fig.~\ref{fig:bmmgmt},
there is significantly more flexibility in the NUHM2 contribution to \gmt\ (solid line). 
However, even
in this case the model is unable to reduce the discrepancy between the
theoretical prediction and the central experimental value much below
the $\Delta \chi^2 \sim 9$ level.
A reduction of the minimum value of the global $\chi^2$ function w.r.t.\ the SM~\cite{mc9} is found
at the level of $\Delta\chi^2 \sim 4.0$, with a best-fit value of $\Delta \left( \frac{g-2}{2} \right) = 3.4 \times 10^{-10}$.
Comparing with the NUHM1 (best-fit value $\Delta \left( \frac{g-2}{2} \right) = 1.0 \times 10^{-10}$), we find a reduction in
the \gmt\ contribution to the global $\chi^2$ function at the best-fit point by $\sim 1.6$, that is largely
compensated by a net increase in the contributions of other observables, including the electroweak
precision measurements. The best-fit value in the CMSSM is $\Delta \left( \frac{g-2}{2} \right) = 2.8 \times 10^{-10}$,
with a total $\chi^2$ higher than in the NUHM2 by 2.5.
As seen in Fig.~\ref{fig:mchi0mchipm}, in the low-mass regions the $\Delta \chi^2$ functions for
$\mneu{1}$ (in the left panel) and $\mcha{1}$ (in the right panel) in
the NUHM2, the NUHM1 and the CMSSM are not very different.
Going to lower mass, as would be needed for a further reduction in the \gmt\ discrepancy, is strongly
penalized by the direct LHC searches for sparticles.

\subsection{Direct Dark Matter Detection}

\begin{figure*}[htb!]
\resizebox{8cm}{!}{\includegraphics{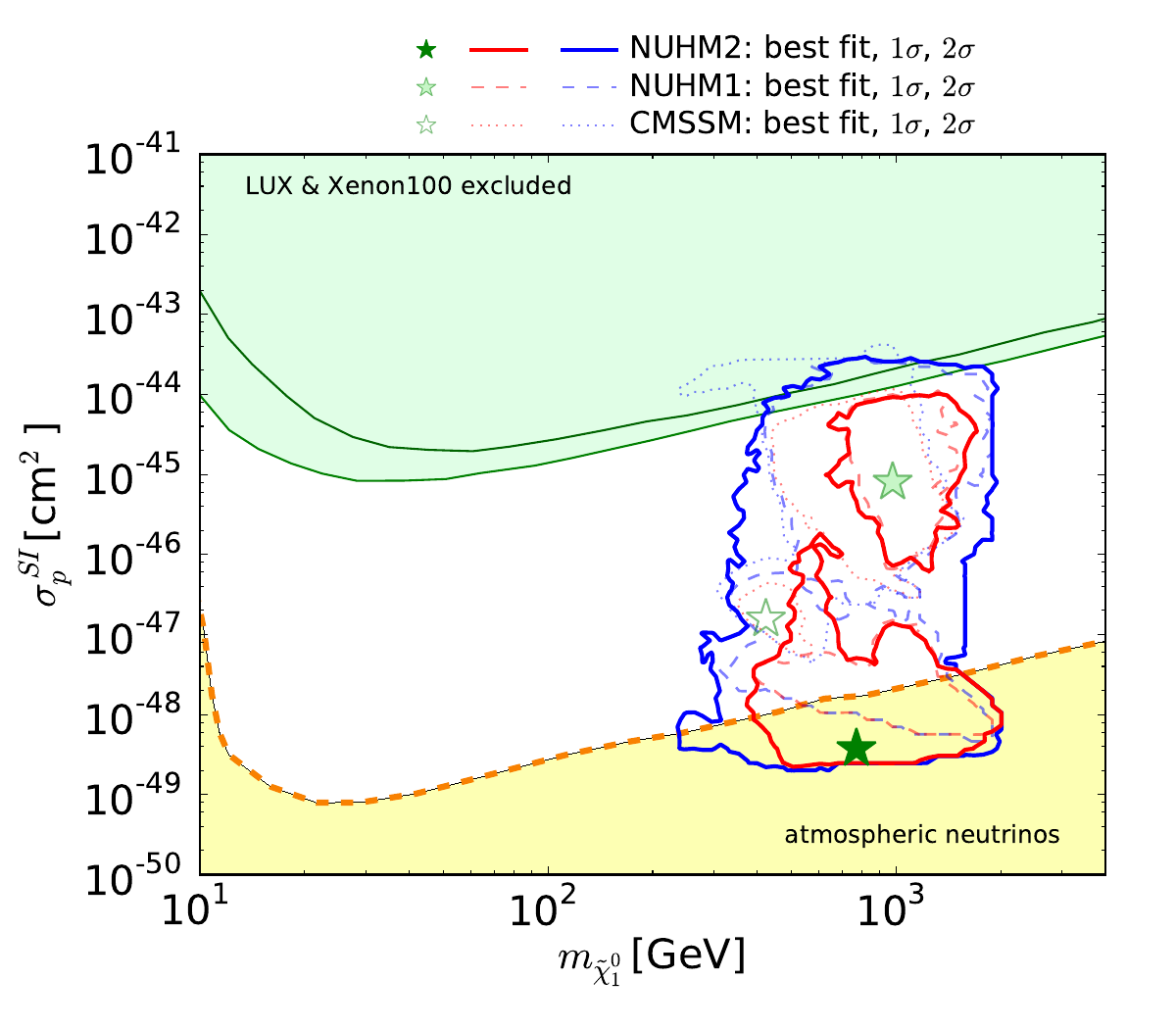}}
\resizebox{8cm}{!}{\includegraphics{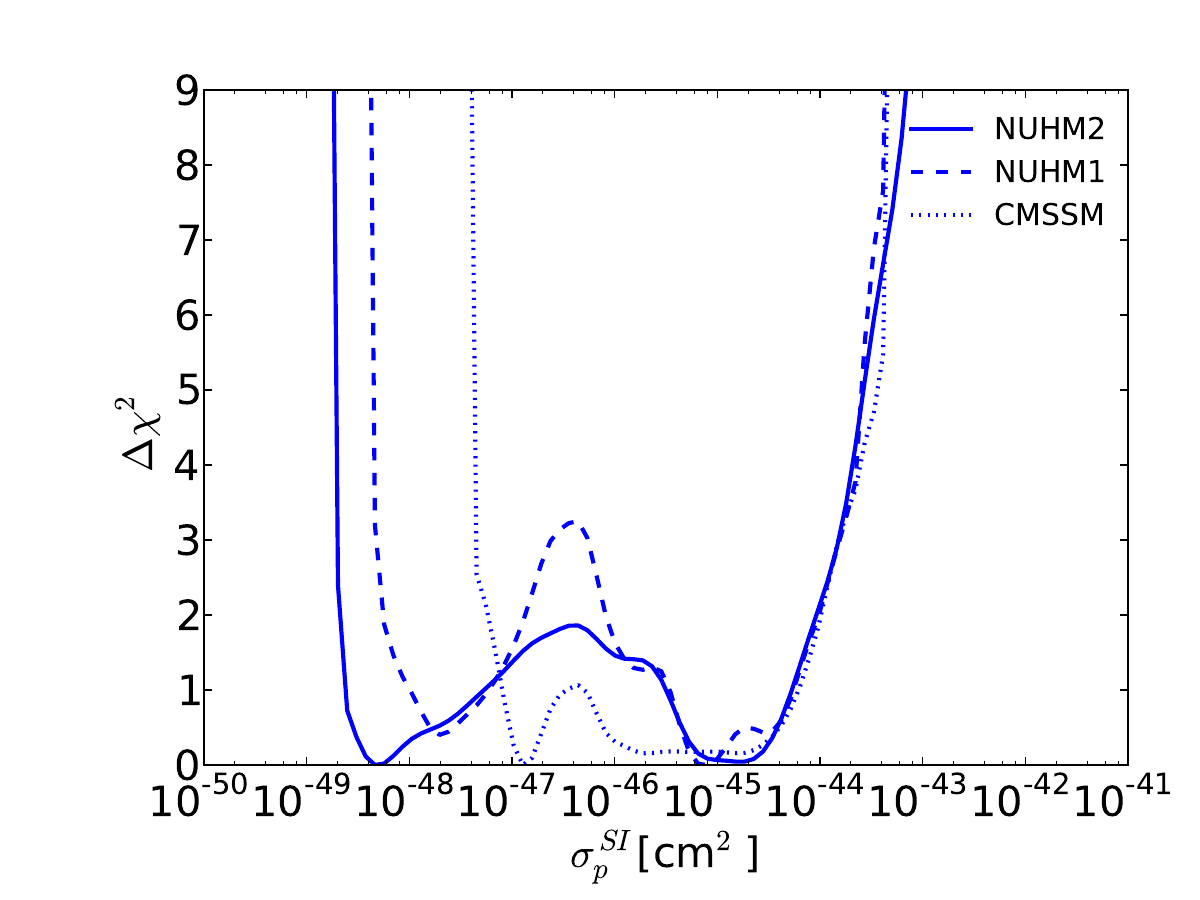}}
\caption{\it Left panel: The $(\mneu{1}, \ssi)$ plane in the NUHM2, with results in the CMSSM and NUHM1
shown for comparison. The
star and contours have the same significations as in Fig.~\protect\ref{fig:m0m12}.
Also shown are the 90\% CL upper limits on \ssi\ from the XENON100~\protect\cite{XENON100}
and LUX~\protect\cite{lux} experiments (green and black lines, respectively),
and the calculated atmospheric neutrino background level from~\protect\cite{nuback}
(orange dashed line). Right panel: The $\Delta \chi^2$
functions for \ssi\ in the CMSSM, NUHM1 and NUHM2.}
\label{fig:mchissi}
\end{figure*}

The left panel of Fig.~\ref{fig:mchissi} displays the $(\mneu{1},\ssi)$ plane,
where \ssi\ is the spin-independent LSP-proton scattering cross-section,
including the best-fit points and the 68\% and 95\% CL contours in the NUHM2,
NUHM1 and CMSSM. Our computation of $\ssi$ follows the procedure
described in~\cite{mc9}, and we have once again adopted for the $\pi$-nucleon $\sigma$ term
the value $\Sigma_{\pi N} = 50 \pm 7$ MeV.
In addition to the model results,
 we also display the 90\% CL upper limits on \ssi\ given by the XENON100 and LUX
 experiments~\cite{XENON100,lux}, and the level of the atmospheric neutrino background~\cite{nuback}.
As we see in the right panel of Fig.~\ref{fig:mchissi}, in the CMSSM the $\Delta \chi^2$
function is relatively flat for $10^{-47}$~cm$^2 \lsim$ \ssi\ $\lsim 10^{-45}$~cm$^2$.
On the other hand, in the case of the NUHM1, values of \ssi\ $\sim 10^{-48}$~cm$^2$ are only
slightly disfavoured relative to the best-fit value of \ssi\ $\sim 10^{-45}$~cm$^2$, with
intermediate values somewhat disfavoured.
In the case of the NUHM2, values of \ssi\ $\sim 4 \times 10^{-49}$~cm$^2$, within the range where the atmospheric
neutrino background dominates, are slightly favoured relative to the range $\ssi\ \sim 10^{-45}$~cm$^2$.
In all the three models, the steep rise in the $\Delta \chi^2$ function at low values of \ssi\
is due to the contribution from Higgs exchange via the small Higgsino component in the $\neu{1}$.

\section{Summary and Conclusions}

In this paper we have presented the results of a frequentist global fit of the NUHM2 model. 
Previous analyses of the CMSSM and NUHM1 models~\cite{mc9} have shown those models 
to be very constrained by available data. One might have wondered whether the extra 
degrees of freedom in the Higgs sector in the NUHM2 scenario would alleviate this tension,
but we found that this was not the case. 

Our fit employed $\sim 4 \times 10^8$ points in the NUHM2 parameter space, and we paid particular
attention to the part of the NUHM2 parameter space where $m_0^2 < 0$. Applying the LHC
constraints on jets + $\ETslash$ to the NUHM1,2 (and especially to 
$m_0^2 < 0$) required an extrapolation from the published
results, which we  previously validated for 7~TeV limits using an implementation of the
{\it {\tt Delphes}} collider detector simulation code set to emulate the ATLAS detector.

The minimum value of $\chi^2/{\rm dof}$ was $32.5/21$, to be
compared with the values 
$\chi^2/{\rm dof} \sim 32.7/22$ and $35.0/23$ found in our
previous analyses of the NUHM1 and 
CMSSM, respectively. We found that ranges of $m_{H_u}^2 < m_{H_d}^2 < m_0^2 < 0$ are favoured.
We find similar tension between \gmt\ and the LHC Higgs and jets +
$\ETslash$ constraints in the NUHM2 as in the NUHM1 and CMSSM.
The best-fit values of $\mgl$ and $\msqr$ in the NUHM2 are $\sim 3 \tev$, with
$\chi^2$ functions that are quite flat for masses $\gsim 2 \tev$. 
The freedom effectively to vary $\mu$ and $\MA$ in the NUHM2 does not suffice to provide a better fit to \gmt\, 
and suggests that if this anomaly persists then some non-universality among the SUSY-breaking scalar masses may be required.

On the one hand, it is encouraging that the results of this NUHM2
analysis are relatively similar to those found previously for the NUHM1
and the CMSSM, suggesting that the type of frequentist analysis
presented here is robust with respect to simple expansions of the CMSSM
parameter space. On the other hand, this analysis suggests that it would
be interesting to study models in which the GUT universality assumptions
are further relaxed, with a corresponding increase in the number of
parameters. Such models may offer the prospect of a significant
reduction in $\chi^2$ if they can relax the tension between \gmt\ and
the LHC constraints. Similarly, models based on a phenomenological definition
of low-energy soft supersymmetry-breaking parameters, variants of the pMSSM~\cite{pMSSM},
may also ameliorate the tension. This may offer another path of extension beyond the well-studied CMSSM, NUHM1 and NUHM2 scenarios.


\subsubsection*{Acknowledgements}

The work of O.B., J.E., S.M., K.A.O. and K.J.dV. is supported in part by
the London Centre for Terauniverse Studies (LCTS), using funding from
the European Research Council 
via the Advanced Investigator Grant 267352. The work of J.E. is also supported in part by STFC
(UK) under the research grant ST/J002798/1. 
The work of S.H. is supported 
in part by CICYT (grant FPA 2013-40715-P) and by the
Spanish MICINN's Consolider-Ingenio 2010 Program under grant MultiDark
CSD2009-00064. The work of K.A.O. is supported in part by DOE grant
DE-SC0011842 at the University of Minnesota. The work of G.W.\ is supported in 
part by the Collaborative Research Center SFB676 of the DFG, ``Particles, Strings and the early Universe", 
and by the European Commission through the ``HiggsTools" Initial Training Network PITN-GA-2012-316704.



\begin{thebibliography}{99}

\bibitem{Ellis:2000ig} 
L. Maiani, {\it Proceedings of the 1979 Gif-sur-Yvette Summer School On Particle Physics},
G. 't Hooft, in {\it Recent Developments in Gauge Theories, Proceedings of the Nato Advanced Study Institute,
Cargese, 1979}, eds. G. Õt Hooft et al., (Plenum Press, NY, 1980);
E.~Witten,
  Phys.\ Lett.\ B {\bf 105} (1981) 267.
  
\bibitem{unify}
J.~Ellis, S.~Kelley and D.V. Nanopoulos,
Phys.\ Lett.\ B {\bf 249} (1990) 441;
Phys.\ Lett.\ B {\bf 260} (1991) 131;
U.~Amaldi, W.~de~Boer and H.~Furstenau, 
Phys.\ Lett.\ B {\bf 260} (1991) 447;
P.~Langacker and M.-x. Luo,
Phys.\ Rev.\ D {\bf 44} (1991) 817;
C.~Giunti, C.~W. Kim and U.~W. Lee,
Mod.\ Phys.\ Lett.\ A {\bf 6} (1991) 1745.

\bibitem{EHNOS} H.~Goldberg,
                Phys.\ Rev.\ Lett.\ {\bf 50} (1983) 1419;
                J.~Ellis, J.~Hagelin, D.~Nanopoulos, K.~Olive and M.~Srednicki,
                Nucl.\ Phys.\ B {\bf 238} (1984) 453.

 \bibitem{ATLAS20}
G.~Aad {\it et al.}  [ATLAS Collaboration],
  arXiv:1405.7875 [hep-ex].
  
\bibitem{CMS20}
S.~Chatrchyan {\it et al.}  [CMS Collaboration],
  JHEP {\bf 1406} (2014) 055
  [arXiv:1402.4770 [hep-ex]].
\bibitem{lhch}
G.~Aad {\it et al.}  [ATLAS Collaboration],
  Phys.\ Lett.\ B {\bf 716} (2012) 1
  [arXiv:1207.7214 [hep-ex]];
   S.~Chatrchyan {\it et al.}  [CMS Collaboration],
  Phys.\ Lett.\ B {\bf 716} (2012) 30 
  [arXiv:1207.7235 [hep-ex]].

 \bibitem{LHCbBsmm}
 R.Aaij {\it et al.}  [LHCb Collaboration],
 Phys.\ Rev.\ Lett.\  {\bf 111} (2013) 101805
 [arXiv:1307.5024 [hep-ex]].

 \bibitem{CMSBsmm}
  S.~Chatrchyan {\it et al.}  [CMS Collaboration],
 Phys.\ Rev.\ Lett.\  {\bf 111} (2013) 101804
 [arXiv:1307.5025 [hep-ex]].

 \bibitem{BsmmComb}
 R.Aaij {\it et al.}  [LHCb and CMS Collaborations],
LHCb-CONF-2013-012, CMS PAS BPH-13-007.

   \bibitem{HK}
  H.~P.~Nilles, Phys.\ Rept.\ {\bf 110} (1984) 1;
H.~E.~Haber and G.~L.~Kane,
  Phys.\ Rept.\  {\bf 117} (1985) 75.

\bibitem{mc9} O.~Buchmueller {\it et al.},
  Eur.\ Phys. J. C {\bf 74} (2014) 2922
  [arXiv:1312.5250 [hep-ph]].

\bibitem{otherRun1}
    T.~Li, J.~A.~Maxin, D.~V.~Nanopoulos and J.~W.~Walker,
  Phys.\ Lett.\ B {\bf 710} (2012) 207
  [arXiv:1112.3024 [hep-ph]];
   M.~J.~Dolan {\it et al.},
    JHEP {\bf 1106} (2011) 095
  [arXiv:1104.0585 [hep-ph]].
    S.~Heinemeyer, O.~Stal and G.~Weiglein,
  Phys.\ Lett.\ B {\bf 710} (2012) 201 
  [arXiv:1112.3026 [hep-ph]];
A.~Arbey, M.~Battaglia, A.~Djouadi, F.~Mahmoudi and J.~Quevillon,
  Phys.\ Lett.\ B {\bf 708} (2012) 162
  [arXiv:1112.3028 [hep-ph]];
   P.~Draper, P.~Meade, M.~Reece and D.~Shih,
  Phys.\ Rev.\ D {\bf 85} (2012) 095007
  [arXiv:1112.3068 [hep-ph]];
  S.~Akula, B.~Altunkaynak, D.~Feldman, P.~Nath and G.~Peim,
  Phys.\ Rev.\ D {\bf 85} (2012) 075001
  [arXiv:1112.3645 [hep-ph]];
M.~Kadastik, K.~Kannike, A.~Racioppi and M.~Raidal,
  JHEP {\bf 1205} (2012) 061
  [arXiv:1112.3647 [hep-ph]];
  C.~Strege {\it et al.},
  JCAP {\bf 1203} (2012) 030 
  [arXiv:1112.4192 [hep-ph]];
  J.~Cao, Z.~Heng, D.~Li and J.~M.~Yang,
  Phys.\ Lett.\ B {\bf 710} (2012) 665
  [arXiv:1112.4391 [hep-ph]];
   L.~Aparicio, D.~G.~Cerdeno and L.~E.~Ibanez,
  JHEP {\bf 1204} (2012) 126
  [arXiv:1202.0822 [hep-ph]]; 
  H.~Baer, V.~Barger and A.~Mustafayev,
  JHEP {\bf 1205} (2012) 091
  [arXiv:1202.4038 [hep-ph]];
   P.~Bechtle {\it et al.},
  JHEP {\bf 1206} (2012) 098 
  [arXiv:1204.4199 [hep-ph]];
Eur.\ Phys.\ J. C {\bf 73} (2013) 2563
 [arXiv:1205.1568 [hep-ph]];
D.~Ghosh, M.~Guchait, S.~Raychaudhuri and D.~Sengupta,
  Phys.\ Rev.\ D {\bf 86} (2012) 055007
  [arXiv:1205.2283 [hep-ph]];
   A.~Fowlie, M.~Kazana, K.~Kowalska, S.~Munir, L.~Roszkowski, E.~M.~Sessolo, S.~Trojanowski and Y.~-L.~S.~Tsai,
  Phys.\ Rev.\ D {\bf 86} (2012) 075010
  [arXiv:1206.0264 [hep-ph]];
   K.~Kowalska {\it et al.}  [BayesFITS Group Collaboration],
  Phys.\ Rev.\ D {\bf 87} (2013) 115010 
  [arXiv:1211.1693 [hep-ph]];
   C.~Strege, G.~Bertone, F.~Feroz, M.~Fornasa, R.~Ruiz de Austri and R.~Trotta,
  JCAP {\bf 1304}, 013 (2013)
  [arXiv:1212.2636 [hep-ph]];
  M.~E.~Cabrera, J.~A.~Casas and R.~R.~de Austri,
  JHEP {\bf 1307} (2013) 182
  [arXiv:1212.4821 [hep-ph]];
  T.~Cohen and J.~G.~Wacker,
  JHEP {\bf 1309} (2013) 061
  [arXiv:1305.2914 [hep-ph]];
S.~Henrot-Versill\'e {\it et al.},
  Phys.\ Rev.\ D {\bf 89}, 055017 (2014)
  [arXiv:1309.6958 [hep-ph]];
  P.~Bechtle {\it et al.},
    PoS EPS-HEP2013 (2013) 31
[arXiv:1310.3045 [hep-ph]];
  L.~Roszkowski, E.~M.~Sessolo and A.~J.~Williams,
  arXiv:1405.4289 [hep-ph].

 \bibitem{previousNUHM2}
 H.~Baer, V.~Barger and A.~Mustafayev,
  arXiv:1112.3017 [hep-ph].

  \bibitem{funnel}  
M.~Drees and M.~M.~Nojiri,
Phys.\ Rev.\ D {\bf 47} (1993) 376 [arXiv:hep-ph/9207234];
  H.~Baer and M.~Brhlik,
Phys.\ Rev.\ D {\bf 53} (1996) 597 [arXiv:hep-ph/9508321];
  Phys.\ Rev.\  D {\bf 57} (1998) 567
  [arXiv:hep-ph/9706509];
   H.~Baer, M.~Brhlik, M.~A.~Diaz, J.~Ferrandis, P.~Mercadante, P.~Quintana and X.~Tata,
    Phys.\ Rev.\  D {\bf 63} (2001) 015007
  [arXiv:hep-ph/0005027];
  J.~R.~Ellis, T.~Falk, G.~Ganis, K.~A.~Olive and M.~Srednicki,
  Phys.\ Lett.\ B {\bf 510} (2001) 236
  [hep-ph/0102098].

\bibitem{cmssm}
 G.~L.~Kane, C.~F.~Kolda, L.~Roszkowski and J.~D.~Wells,
  Phys.\ Rev.\  D {\bf 49} (1994) 6173
  [arXiv:hep-ph/9312272];
  J.~R.~Ellis, T.~Falk, K.~A.~Olive and M.~Schmitt,
Phys.\ Lett.\ B {\bf 388} (1996) 97
[arXiv:hep-ph/9607292];
Phys.\ Lett.\ B {\bf 413} (1997) 355
[arXiv:hep-ph/9705444];
J.~R.~Ellis, T.~Falk, G.~Ganis, K.~A.~Olive and M.~Schmitt,
Phys.\ Rev.\ D {\bf 58} (1998) 095002
[arXiv:hep-ph/9801445];
V.~D.~Barger and C.~Kao,
Phys.\ Rev.\ D {\bf 57} (1998) 3131
[arXiv:hep-ph/9704403];
J.~R.~Ellis, T.~Falk, G.~Ganis and K.~A.~Olive,
Phys.\ Rev.\ D {\bf 62} (2000) 075010
[arXiv:hep-ph/0004169];
L.~Roszkowski, R.~Ruiz de Austri and T.~Nihei,
JHEP {\bf 0108} (2001) 024
[arXiv:hep-ph/0106334];
  A.~Djouadi, M.~Drees and J.~L.~Kneur,
JHEP {\bf 0108} (2001) 055
[arXiv:hep-ph/0107316];
U.~Chattopadhyay, A.~Corsetti and P.~Nath,
Phys.\ Rev.\ D {\bf 66} (2002) 035003
[arXiv:hep-ph/0201001];
J.~R.~Ellis, K.~A.~Olive and Y.~Santoso,
New Jour.\ Phys.\  {\bf 4} (2002) 32
[arXiv:hep-ph/0202110];
H.~Baer, C.~Balazs, A.~Belyaev, J.~K.~Mizukoshi, X.~Tata and Y.~Wang,
JHEP {\bf 0207} (2002) 050
[arXiv:hep-ph/0205325];
R.~Arnowitt and B.~Dutta,
arXiv:hep-ph/0211417.

\bibitem{AbdusSalam:2011fc}
  S.~S.~AbdusSalam,  {\it et al.},
  Eur.\ Phys.\ J.\ C {\bf 71} (2011) 1835 
  [arXiv:1109.3859 [hep-ph]].

   \bibitem{nuhm1}
H.~Baer, A.~Mustafayev, S.~Profumo, A.~Belyaev and X.~Tata,
  Phys.\ Rev.\  D {\bf 71} (2005) 095008
  [arXiv:hep-ph/0412059];
            H.~Baer, A.~Mustafayev, S.~Profumo, A.~Belyaev and X.~Tata,
               {\em JHEP} {\bf 0507} (2005) 065, 
               hep-ph/0504001;
  J.~R.~Ellis, K.~A.~Olive and P.~Sandick,
  Phys.\ Rev.\  D {\bf 78} (2008) 075012
  [arXiv:0805.2343 [hep-ph]];
 J.~Ellis, F.~Luo, K.~A.~Olive and P.~Sandick,
  Eur.\ Phys.\ J.\ C {\bf 73} (2013) 2403
  [arXiv:1212.4476 [hep-ph]].

   \bibitem{newBNL} 
 G.~Bennett et al.\ [The Muon g-2 Collaboration],
 {\it Phys. Rev. Lett.} {\bf 92} (2004) 161802, 
 [arXiv:hep-ex/0401008]; and
  {\em Phys.\ Rev.} {\bf D 73} (2006) 072003
  [arXiv:hep-ex/0602035].

   \bibitem{g-2}
  D.~Stockinger,
  J.\ Phys.\ G {\bf 34} (2007) R45
  [arXiv:hep-ph/0609168];
    J.~Miller, E.~de~Rafael and B.~Roberts,
   {\em Rept.\ Prog.\ Phys.} {\bf 70} (2007) 795
   [arXiv:hep-ph/0703049];
    J.~Prades, E.~de Rafael and A.~Vainshtein,
  arXiv:0901.0306 [hep-ph];
    F.~Jegerlehner and A.~Nyffeler,
  Phys.\ Rept.\  {\bf 477}, 1 (2009)
  [arXiv:0902.3360 [hep-ph]];
   M.~Davier, A.~Hoecker, B.~Malaescu, C.~Z.~Yuan and Z.~Zhang,
  Eur.\ Phys.\ J.\  C {\bf 66}, 1 (2010)
  [arXiv:0908.4300 [hep-ph]].
  J.~Prades,
  Acta Phys.\ Polon.\ Supp.\  {\bf 3}, 75 (2010)
  [arXiv:0909.2546 [hep-ph]];
    T.~Teubner, K.~Hagiwara, R.~Liao, A.~D.~Martin and D.~Nomura,
  arXiv:1001.5401 [hep-ph];
  M.~Davier, A.~Hoecker, B.~Malaescu and Z.~Zhang,
  Eur.\ Phys.\ J.\  C {\bf 71} (2011) 1515
  [arXiv:1010.4180 [hep-ph]].

    \bibitem{Jegerlehner}
F.~Jegerlehner and R.~Szafron,
  Eur.\ Phys.\ J.\  C {\bf 71} (2011) 1632
  [arXiv:1101.2872 [hep-ph]];
  M.~Benayoun, P.~David, L.~DelBuono and F.~Jegerlehner,
  Eur.\ Phys.\ J.\ C {\bf 73} (2013) 2453
  [arXiv:1210.7184 [hep-ph]].

  \bibitem{mc7}
O.~Buchmueller {\it et al.},
Eur.\ Phys.\ J.\ C {\bf 72} (2012) 1878
[arXiv:1110.3568 [hep-ph]].

\bibitem{nuhm2}
J.~Ellis, K.~Olive and Y.~Santoso,
Phys.\ Lett.\  B~{\bf 539} (2002) 107
[arXiv:hep-ph/0204192];
J.~R.~Ellis, T.~Falk, K.~A.~Olive and Y.~Santoso,
Nucl.\ Phys.\ B {\bf 652} (2003) 259
[arXiv:hep-ph/0210205].

\bibitem{Feroz:2008xx}
  F.~Feroz, M.~P.~Hobson and M.~Bridges,
  Mon.\ Not.\ Roy.\ Astron.\ Soc.\  {\bf 398} (2009) 1601
  [arXiv:0809.3437 [astro-ph]].

\bibitem{HFAG}
The Heavy Flavor Averaging Group, D.~Asner {\it et al.}, 
arXiv:1010.1589 [hep-ex], with updates available at 
{\tt http://www.slac.stanford.edu/xorg/}
{\tt hfag/osc/end$\underline{~}$2009.}

\bibitem{EWWG}
LEP Electroweak Working Group, {\tt http://lepewwg.web.cern.ch/LEPEWWG/}.

  \bibitem{XENON100}
  E.~Aprile {\it et al.}  [XENON100 Collaboration],
  Phys.\ Rev.\ Lett.\  {\bf 107} (2011) 131302
  [arXiv:1104.2549 [astro-ph.CO]].

  \bibitem{lux}
   D.~S.~Akerib {\it et al.}  [LUX Collaboration],
  Phys.\ Rev.\ Lett.\  {\bf 112}, 091303 (2014)
  [arXiv:1310.8214 [astro-ph.CO]].

\bibitem{Craig:2012xp}
  N.~Craig, S.~Knapen, D.~Shih and Y.~Zhao,
  JHEP {\bf 1303} (2013) 154
  [arXiv:1206.4086 [hep-ph]].

  \bibitem{EGLOS}
J.~R.~Ellis, J.~Giedt, O.~Lebedev, K.~Olive and M.~Srednicki,
  Phys.\ Rev.\ D {\bf 78} (2008) 075006
  [arXiv:0806.3648 [hep-ph]].

\bibitem{mc8} O.~Buchmueller {\it et al.},
  Eur.\ Phys.\ J.\ C {\bf 72} (2012) 2243
  [arXiv:1207.7315].

\bibitem{mcweb}
For more information and updates, please see {\tt http://cern.ch/mastercode/}.

\bibitem{Svenetal}
  S.~Heinemeyer {\it et al.}, 
  JHEP {\bf 0608} (2006) 052
  [arXiv:hep-ph/0604147];
  S.~Heinemeyer, W.~Hollik, A.~M.~Weber and G.~Weiglein,
  JHEP {\bf 0804} (2008) 039
  [arXiv:0710.2972 [hep-ph]].

\bibitem{Allanach:2001kg}
  B.~C.~Allanach,
  Comput.\ Phys.\ Commun.\  {\bf 143} (2002) 305
  [arXiv:hep-ph/0104145].

\bibitem{FeynHiggs}
 G.~Degrassi, S.~Heinemeyer, W.~Hollik, P.~Slavich and G.~Weiglein,
  Eur.\ Phys.\ J.\ C {\bf 28} (2003) 133
  [arXiv:hep-ph/0212020];
   S.~Heinemeyer, W.~Hollik and G.~Weiglein,
  Eur.\ Phys.\ J.\ C {\bf 9} (1999) 343
  [arXiv:hep-ph/9812472];
  S.~Heinemeyer, W.~Hollik and G.~Weiglein,
  Comput.\ Phys.\ Commun.\  {\bf 124} (2000) 76
  [arXiv:hep-ph/9812320];
   M.~Frank {\it et al.}, 
  JHEP {\bf 0702} (2007) 047
  [arXiv:hep-ph/0611326];
  See {\tt http://www.feynhiggs.de}~.

\bibitem{Mh-logresum} T.~Hahn, S.~Heinemeyer, W.~Hollik, H.~Rzehak and
  G.~Weiglein, 
  Phys.\ Rev.\ Lett.\  {\bf 112} (2014) 141801
  [arXiv:1312.4937 [hep-ph]].

\bibitem{SuFla}
 G.~Isidori and P.~Paradisi,
  Phys.\ Lett.\ B {\bf 639} (2006) 499
  [arXiv:hep-ph/0605012];
  G.~Isidori, F.~Mescia, P.~Paradisi and D.~Temes,
  Phys.\ Rev.\  D {\bf 75} (2007) 115019
  [arXiv:hep-ph/0703035], and references therein.

\bibitem{SuperIso}
F.~Mahmoudi,
  Comput.\ Phys.\ Commun.\  {\bf 178} (2008) 745
  [arXiv:0710.2067 [hep-ph]]; 
  Comput.\ Phys.\ Commun.\  {\bf 180} (2009) 1579
  [arXiv:0808.3144 [hep-ph]];
  D.~Eriksson, F.~Mahmoudi and O.~Stal,
  JHEP {\bf 0811} (2008) 035
  [arXiv:0808.3551 [hep-ph]].

\bibitem{MicroMegas}
  G.~Belanger, F.~Boudjema, A.~Pukhov and A.~Semenov,
  Comput.\ Phys.\ Commun.\  {\bf 176} (2007) 367
  [arXiv:hep-ph/0607059];
  Comput.\ Phys.\ Commun.\  {\bf 149} (2002) 103
  [arXiv:hep-ph/0112278];
  Comput.\ Phys.\ Commun.\  {\bf 174} (2006) 577
  [arXiv:hep-ph/0405253].

\bibitem{SSARD}  Information about this code is available from K.~A.~Olive: it contains important contributions 
from T.~Falk, A.~Ferstl, G.~Ganis, A.~Mustafayev, J.~McDonald, F. Luo, K.~A.~Olive, P.~Sandick, Y.~Santoso, V. Spanos, and M.~Srednicki. 

\bibitem{SLHA}
P.~Skands {\it et al.},
  JHEP {\bf 0407} (2004) 036
  [arXiv:hep-ph/0311123];
  B.~Allanach {\it et al.},
  Comput.\ Phys.\ Commun.\  {\bf 180} (2009) 8
  [arXiv:0801.0045 [hep-ph]].

\bibitem{mc8.5} O.~Buchmueller {\it et al.},
  Eur.\ Phys.\ J.\ C {\bf 74} (2014) 2809
  [arXiv:1312.5233 [hep-ph]].

  \bibitem{Delphes}
For a description of {\tt Delphes}, 
written by S.~Ovyn and X.~Rouby, see
{\tt http://www.fynu.ucl.ac.be/users/s.ovyn/} {\tt Delphes/index.html}.

  \bibitem{m2neg}
 J.~L.~Feng, A.~Rajaraman and B.~T.~Smith,
  Phys.\ Rev.\  D {\bf 74} (2006) 015013 
  [arXiv:hep-ph/0512172];
 A.~Rajaraman and B.~T.~Smith,
  Phys.\ Rev.\  D {\bf 75} (2007) 115015 
  [arXiv:hep-ph/0612235].
  
    \bibitem{cms0l-aT} V.~Khachatryan {\it et al.}  [CMS Collaboration],
  Phys.\ Lett.\ B {\bf 698} (2011) 196
  [arXiv:1101.1628 [hep-ex]].

   \bibitem{ATLASindependent}
ATLAS Collaboration, \\ {\tt https://cdsweb.cern.ch/record/1432199/\\
files/ATLAS-CONF-2012-033.pdf.}

  \bibitem{ATLASHA}
ATLAS Collaboration, \\{\tt https://cds.cern.ch/record/1744694/} {\tt files/ATLAS-CONF-2014-049.pdf}.
See also
V.~Khachatryan {\it et al.}  [ CMS Collaboration],
  arXiv:1408.3316 [hep-ex].

\bibitem{Martin:1993zk}
S.~P.~Martin and M.~T.~Vaughn,
Phys.\ Rev.\ D {\bf 50} (1994) 2282
[arXiv:hep-ph/9311340].

\bibitem{Sterms}
K.~Inoue, A.~Kakuto, H.~Komatsu and S.~Takeshita,
Prog.\ Theor.\ Phys.\  {\bf 68} (1982) 927
[Erratum-ibid.\  {\bf 70} (1983) 330];
T.~Falk,
Phys.\ Lett.\ B {\bf 456} (1999) 171
[arXiv:hep-ph/9902352].

\bibitem{stau-co} J. Ellis, T. Falk, and K.A. Olive, Phys. Lett.  {\bf B444} (1998) 367
[arXiv:hep-ph/9810360];
J. Ellis, T. Falk, K.A. Olive, and M. Srednicki, {\it Astr. Part. Phys.}
{\bf 13} (2000) 181
[Erratum-ibid.\  {\bf 15} (2001) 413]
[arXiv:hep-ph/9905481];
R.~Arnowitt, B.~Dutta and Y.~Santoso,
Nucl.\ Phys.\ B {\bf 606} (2001) 59
[arXiv:hep-ph/0102181];
M.~E.~G\'omez, G.~Lazarides and C.~Pallis,
Phys. Rev. D {\bf D61} (2000) 123512
[arXiv:hep-ph/9907261];
  Phys.\ Lett. {\bf B487} (2000) 313
[arXiv:hep-ph/0004028];
  Nucl. Phys. B {\bf B638} (2002) 165
[arXiv:hep-ph/0203131];
T.~Nihei, L.~Roszkowski and R.~Ruiz de Austri,
  JHEP {\bf 0207} (2002) 024
[arXiv:hep-ph/0206266];
M.~Citron, J.~Ellis, F.~Luo, J.~Marrouche, K.~A.~Olive and K.~J.~de Vries,
  Phys.\ Rev.\ D {\bf 87}, 036012 (2013)
  [arXiv:1212.2886 [hep-ph]].
  
    \bibitem{edsjo}
       J.~Edsjo, M.~Schelke, P.~Ullio and P.~Gondolo,
  JCAP {\bf 0304}, 001 (2003)
  [hep-ph/0301106].
  
  \bibitem{coann}
  S.~Mizuta and M.~Yamaguchi,
Phys.\ Lett.\ B {\bf 298} (1993) 120
[arXiv:hep-ph/9208251];
  J.~Edsjo and P.~Gondolo,
  Phys.\ Rev.\ D {\bf 56}, 1879 (1997)
  [hep-ph/9704361];
   H.~Baer, C.~Balazs and A.~Belyaev,
  JHEP {\bf 0203}, 042 (2002)
  [hep-ph/0202076];
    A.~Birkedal-Hansen and E.~h.~Jeong,
  JHEP {\bf 0302}, 047 (2003)
  [hep-ph/0210041].
    


 \bibitem{stop}  
  C.~Boehm, A.~Djouadi and M.~Drees,
  Phys.\ Rev.\  D {\bf 62}, 035012 (2000)
  [arXiv:hep-ph/9911496]; 
  J.~R.~Ellis, K.~A.~Olive and Y.~Santoso,
  Astropart.\ Phys.\  {\bf 18}, 395 (2003)
  [arXiv:hep-ph/0112113];
       J.~L.~Diaz-Cruz, J.~R.~Ellis, K.~A.~Olive and Y.~Santoso,
  JHEP {\bf 0705}, 003 (2007)
  [arXiv:hep-ph/0701229];
  I.~Gogoladze, S.~Raza and Q.~Shafi,
  Phys.\ Lett.\ B {\bf 706}, 345 (2012)
  [arXiv:1104.3566 [hep-ph]];
   M.~A.~Ajaib, T.~Li and Q.~Shafi,
  Phys.\ Rev.\ D {\bf 85}, 055021 (2012)
  [arXiv:1111.4467 [hep-ph]];
J.~Ellis, K.~A.~Olive and J.~Zheng,
  arXiv:1404.5571 [hep-ph].

\bibitem{fp}
  J.~L.~Feng, K.~T.~Matchev and T.~Moroi,
  Phys.\ Rev.\ Lett.\  {\bf 84} (2000) 2322
  [arXiv:hep-ph/9908309];
  Phys.\ Rev.\  D {\bf 61} (2000) 075005
  [arXiv:hep-ph/9909334]; 
  J.~L.~Feng, K.~T.~Matchev and F.~Wilczek,
  Phys.\ Lett.\  B {\bf 482} (2000) 388
  [arXiv:hep-ph/0004043].
  
\bibitem{nuback}
P.~Cushman {\it et al.},
arXiv:1310.8327 [hep-ex].

\bibitem{pMSSM}
See, for example,
C.~F.~Berger, J.~S.~Gainer, J.~L.~Hewett and T.~G.~Rizzo,
  JHEP {\bf 0902}, 023 (2009)
  [arXiv:0812.0980 [hep-ph]];
S.~S.~AbdusSalam, B.~C.~Allanach, F.~Quevedo, F.~Feroz and M.~Hobson,
  Phys.\ Rev.\ D {\bf 81}, 095012 (2010)
  [arXiv:0904.2548 [hep-ph]];
  J.~A.~Conley, J.~S.~Gainer, J.~L.~Hewett, M.~P.~Le and T.~G.~Rizzo,
  Eur.\ Phys.\ J.\ C {\bf 71}, 1697 (2011)
  [arXiv:1009.2539 [hep-ph]];
  J.~A.~Conley, J.~S.~Gainer, J.~L.~Hewett, M.~P.~Le and T.~G.~Rizzo,
  [arXiv:1103.1697 [hep-ph]];
  S.~Sekmen, S.~Kraml, J.~Lykken, F.~Moortgat, S.~Padhi, L.~Pape, M.~Pierini and H.~B.~Prosper {\it et al.},
  JHEP {\bf 1202} (2012) 075
  [arXiv:1109.5119 [hep-ph]];
  A.~Arbey, M.~Battaglia and F.~Mahmoudi,
  Eur.\ Phys.\ J.\ C {\bf 72} (2012) 1847
  [arXiv:1110.3726 [hep-ph]];
  A.~Arbey, M.~Battaglia, A.~Djouadi and F.~Mahmoudi,
  Phys.\ Lett.\ B {\bf 720} (2013) 153
  [arXiv:1211.4004 [hep-ph]];
  M.~W.~Cahill-Rowley, J.~L.~Hewett, A.~Ismail and T.~G.~Rizzo,
  Phys.\ Rev.\ D {\bf 88} (2013) 3,  035002
  [arXiv:1211.1981 [hep-ph]].

\end{thebibliography}
\end{document}